\documentclass[9pt,twocolumn,twoside]{gsajnl}

\usepackage[utf8]{inputenc}
\usepackage[T1]{fontenc}
\usepackage{amsfonts}
\usepackage{amssymb}
\usepackage{pifont}
\usepackage{amsmath}
\usepackage[super]{natbib}
\usepackage{hyperref}
\hypersetup{
    colorlinks=true,
    linkcolor=blue,
    filecolor=magenta,      
    urlcolor=cyan,
    pdftitle={Overleaf Example},
    pdfpagemode=FullScreen,
    }
\usepackage[capitalise]{cleveref}
\usepackage{placeins}
\usepackage{array}

\usepackage{graphicx}
\usepackage{booktabs}

\usepackage{caption}        
\usepackage{subcaption}     
\usepackage{multirow}
\usepackage{xcolor}
\usepackage{enumitem}

\usepackage{microtype}
\usepackage{contour}
\usepackage{adjustbox}

\renewenvironment{thebibliography}[1]{
  \begin{oldthebibliography}{#1}
    \setlength{\itemsep}{0em}
    \setlength{\parskip}{0em}
}
{
  \end{oldthebibliography}
}

\title{OCTolyzer: Fully automatic toolkit for segmentation and feature extracting in optical coherence tomography and scanning laser ophthalmoscopy data}
\runningtitle{OCTolyzer: Fully Automatic OCT+SLO Image Analysis}
\runningauthor{Burke, et al.}

\author[1,2,*]{Jamie Burke}
\author[3,4]{Justin Engelmann}
\author[2]{Samuel Gibbon}
\author[5]{Charlene Hamid}
\author[6]{Diana Moukaddem}
\author[7]{Dan Pugh}
\author[7]{Tariq Farrah}
\author[6]{Niall Strang}
\author[7]{Neeraj Dhaun}
\author[5,8]{Tom MacGillivray}
\author[1,\textsuperscript{\textdagger}]{Stuart King}
\author[8,9,\textsuperscript{\textdagger}]{Ian J.C. MacCormick}

\affil[1]{School of Mathematics, University of Edinburgh, Edinburgh, UK}
\affil[2]{Robert O Curle Ophthalmology Suite, Institute for Regeneration and Repair, University of Edinburgh, UK}
\affil[3]{School of Informatics, University of Edinburgh, Edinburgh, UK}
\affil[4]{Centre for Medical Informatics, University of Edinburgh, Edinburgh, UK}
\affil[5]{Clinical Research Facility and Imaging, University of Edinburgh, Edinburgh, UK}
\affil[6]{Department of Vision Sciences, Glasgow Caledonian University, Glasgow, UK}
\affil[7]{British Heart Foundation Centre for Cardiovascular Science, University of Edinburgh, Edinburgh, UK}
\affil[8]{Centre for Clinical Brain Sciences, University of Edinburgh, Edinburgh, UK }
\affil[9]{Institute for Adaptive and Neural Computation, School of Informatics, University of Edinburgh, Edinburgh, UK}

\correspondingauthoraffiliation[$\ast$]{Corresponding and lead author; \\ Email address: \url{Jamie.Burke@ed.ac.uk}\\ \textsuperscript{\textdagger}Equal contribution }

\begin{abstract}
Optical coherence tomography (OCT) and scanning laser ophthalmoscopy (SLO) of the eye has become essential to ophthalmology and the emerging field of oculomics, thus requiring a need for transparent, reproducible, and rapid analysis of this data for clinical research and the wider research community. Here, we introduce OCTolyzer, the first open-source toolkit for retinochoroidal analysis in OCT/SLO data. It features two analysis suites for OCT and SLO data, facilitating deep learning-based anatomical segmentation and feature extraction of the cross-sectional retinal and choroidal layers and en face retinal vessels. We describe OCTolyzer and evaluate the reproducibility of its OCT choroid analysis. At the population level, metrics for choroid region thickness were highly reproducible, with a mean absolute error (MAE)/Pearson correlation for macular volume choroid thickness (CT) of 6.7$\mu$m/0.99, macular B-scan CT of 11.6$\mu$m/0.99, and peripapillary CT of 5.0$\mu$m/0.99. Macular choroid vascular index (CVI) also showed strong reproducibility, with MAE/Pearson for volume CVI yielding 0.0271/0.97 and B-scan CVI 0.0130/0.91. At the eye level, measurement noise for regional and vessel metrics was below 5\% and 20\% of the population's variability, respectively. Outliers were caused by poor-quality B-scans with thick choroids and invisible choroid-sclera boundary. Processing times on a laptop CPU were under three seconds for macular/peripapillary B-scans and 85 seconds for volume scans. OCTolyzer can convert OCT/SLO data into reproducible and clinically meaningful retinochoroidal features and will improve the standardisation of ocular measurements in OCT/SLO image analysis, requiring no specialised training or proprietary software to be used. OCTolyzer is freely available here: \url{https://github.com/jaburke166/OCTolyzer}.

\end{abstract}

\begin{document}

\maketitle     

\section{Introduction}
Optical coherence tomography (OCT) of the retina has become essential to clinical and computational ophthalmology, and is becoming routinely collected by many community opticians and clinics \cite{ly2017self, jindal2019impact, chopra2021optical, song2021review}. OCT systems provide a cross-sectional visualisation of the retinal and choroidal layer, and often contains an en face scanning laser ophthalmoscopy (SLO) image of the en face retinal vessels, and there is widespread anticipation that OCT/SLO-derived features may give insights into systemic health and disease \cite{wagner2020insights}. As the accessibility of OCT systems improves in terms of cost, size and availability within and outside the clinic, there is a need for transparent, reproducible and rapid analysis of OCT and SLO data in clinical research in ophthalmology as well as in the emerging field of oculomics \cite{wagner2020insights}. 

OCT uses low coherence interferometry to collect accurate depth (axial) and intensity information from the hyper-reflectivity of retinal tissue at the micron level \cite{salmon2018axial}. The confocal, infrared reflectance scanning laser ophthalmoscopy (SLO) image often also acquired is used as a localiser to position the OCT beam at the back of the eye (Supplementary \cref{suppfig:heyex_demo}). SLO images show the en face, superficial retinal vessels along the inner surface of the retina and is very similar to colour fundus photography (CFP) but with a restricted field of view of 30 degrees (approximately 9 mm$^2$).

The OCT/SLO system together capture the retina and choroid and permit a unique assessment of the microvasculature which play a critical role in maintaining eye health. Interestingly, there is increasing evidence to suggest that the retinal and choroidal circulations correspond with microvascular changes in structure and function of vital organs like the brain and kidney \cite{balmforth2016chorioretinal, farrah2020eye, farrah2023choroidal, gharbiya2014choroidal, ma2022longitudinal, kundu2023longitudinal}. These observations contribute to the nascent field of oculomics, the relationship between the ocular system and systemic disease \cite{wagner2020insights}.

OCT systems have often been focused on imaging the retinal layers, and the choroid has traditionally received lesser attention since older OCT technology was not very effective at imaging structures posterior to the retinal pigment epithelium. However, recent advances have improved images of the choroid, enabling cross-sectional visualisation of retinochoroidal structures in OCT \cite{spaide2008enhanced}. 

The localiser SLO image of the OCT system has also not received much attention because its role has been mainly to orientate the observer to locations within the OCT cross-sectional image stack. However, the en face SLO has several valuable characteristics over CFP: its confocal imaging method produces greater contrast of the vessels, optic disc and fovea \cite{hoyt2012pediatric}, and the interferometry of the OCT system has the potential to visualise the en face retinal vessels in the transverse direction more accurately \cite{scoles2022inaccurate, burke2024sloctolyzer} --- given known biometric factors of the eye. This is a crucial step in generating clinically meaningful, physical measurements of the retina. In contrast, for CFP, comprising a microscope attached to a camera with a flash, heuristic approaches are commonly used to obtain physical measurements, such as optic disc area normalisation \cite{schanner2023impact}.

Anatomical annotation of the retina and choroid on OCT/SLO is prohibitively expensive in terms of time and labour, and is prone to human error. Therefore, there has been a wealth of research into automatic segmentation methods. The majority have focused on retinal OCT layer segmentation, with more than 60 studies examined in a recent comprehensive review \cite{viedma2022deep}. There has been less focus on choroid segmentation methods \cite{mazzaferri2017open, khaing2021choroidnet, betzler2022choroidal, wang2023choroidal, xuan2023deep, arian2023automatic, wang2023choroidal, wen2024transformer} and far fewer methods developed for retinal vessel segmentation in SLO \cite{xu2008retinal, pellegrini2014blood, kromer2016automated, meyer2017deep}. 

Additionally, most of these previous methods are either closed-source \cite{khaing2021choroidnet,
wang2023choroidal, wang2023choroidal, wen2024transformer, xu2008retinal, pellegrini2014blood, kromer2016automated, meyer2017deep}, require permission \cite{betzler2022choroidal} or are not easy-to-use for the general researcher \cite{mazzaferri2017open, xuan2023deep, arian2023automatic}. It is also rare for computational methods to be released with accompanying code for feature extraction, visualisation, quality inspection and batch processing \cite{mazzaferri2017open, xuan2023deep, ss2019octtools, brandt2018octmarker}. There has been increasing demand for easy-to-use, open-source software \cite{perez2011vampire, zhou2022automorph, engelmann2022robust, engelmann2023choroidalyzer, burke2023open} which facilitates standardised measurement of the retina and choroid, and is critical for ensuring the validity, reliability, and comparability of data across different studies \cite{xie2021evaluation}. This software currently does not exist for combined OCT/SLO datasets.

Accordingly, we have developed a fully automatic analysis toolkit, OCTolyzer, for segmentation and feature extraction of cross-sectional retinal and choroidal OCT images and the accompanying en face SLO localiser image. OCTolyzer is designed to equip the general researcher, who may not have a technical background or specialist training in image analysis, with the means of analysing their own clinical OCT/SLO data in a standardised, reliable and reproducible manner. 

\section{Methods}

\begin{figure*}[t]
    \centering
    \includegraphics[width=0.89\textwidth]{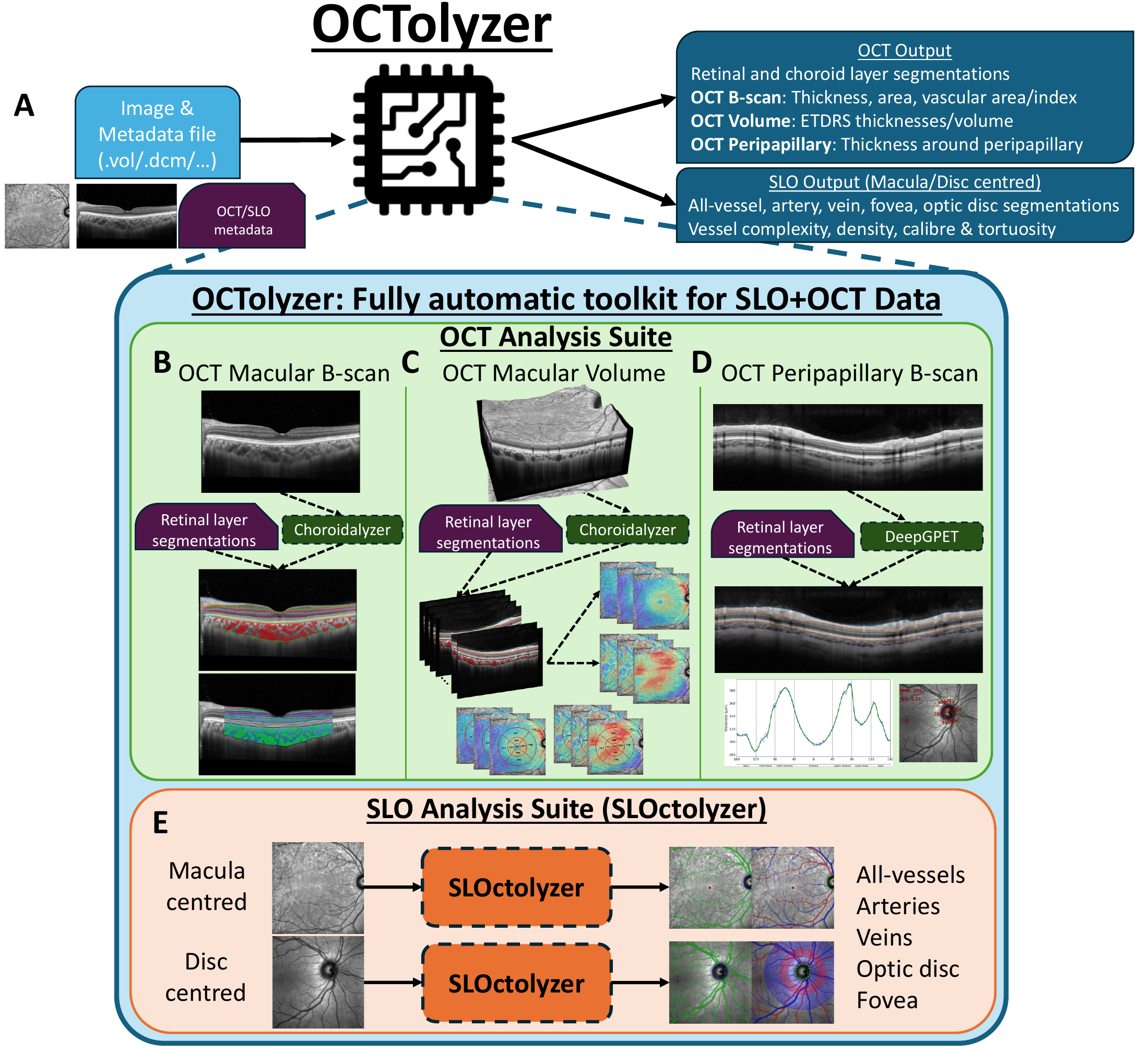}
    \caption{OCTolyzer’s pipeline. (A) Input: OCT (and optional SLO) image data with necessary metadata. (B -- D) OCT analysis suite for single/radial macular B-scans (B), macular volume scans (C), or peripapillary B-scans (D). (E) SLO analysis suite (SLOctolyzer \cite{burke2024sloctolyzer}) for macula- or disc-centred localiser images.}
    \label{fig:pipeline}
\end{figure*}

\cref{fig:pipeline} describes the core elements of OCTolyzer's analysis pipeline. OCTolyzer contains two analysis suites, for the OCT and the corresponding localiser SLO data. OCTolyzer is freely available here: \url{https://github.com/jaburke166/OCTolyzer}.

\subsection{Data}

\subsubsection{Segmentation model populations}
OCTolyzer's segmentation models have been previously published \cite{burke2024sloctolyzer, burke2023open, engelmann2023choroidalyzer}, and the data used to construct them have already been described in their respective studies. The majority of this data were from cohorts related to systemic health and normative data. 

For completeness and brevity, Supplementary \cref{supptab:pop_choroidalyzer} and \cref{supptab:pop_deepgpet}, with supporting text, describe the image and population characteristics for the two deep learning models OCTolyzer uses for OCT choroid segmentation. Additionally, Supplementary \cref{supptab:pop_sloctolyzer} describes the image characteristics of the five cohorts used to build the segmentation models for OCTolyzer's SLO analysis suite. All studies/cohorts \cite{dhaun2014optical, pearson2022multi, hatamizadeh2022ravir, cameron2016modulation, ritchie2012prevent, ritchie2013prevent, chen2022longitudinal, kearns2022futurems} adhered to the Declaration of Helsinki, received relevant ethical approval, and informed consent from all subjects was obtained in all cases from the host institution.

\subsubsection{Reproducibility populations}
The primary analysis in this study assesses the reproducibility of OCTolyzer's choroid segmentation models across three cohorts: i-Test \cite{dhaun2014optical}, Glasgow Caledonian University Topcon (GCU Topcon) \cite{moukaddem2022comparison}, and Diurnal Variation for Chronic Kidney Disease (DVCKD) \cite{dhaun2014optical}. The analysis included all eyes with repeated data (120 eyes from 60 i-Test participants, 33 eyes from 21 GCU Topcon participants, and 22 eyes from 22 DVCKD participants). Core image and population characteristics are detailed in \cref{tab:rep_table}.

\begin{table*}[t]
\centering
{\small\scalebox{0.9}{\centerline{\begin{tabular}{p{3.5cm}p{4.5cm}p{4.5cm}p{4.5cm}}
\toprule
\multirow{2}{*}{} & \multicolumn{3}{c}{Study} \\
\cmidrule(l){2-4}
 & i-Test \cite{dhaun2014optical} & GCU Topcon \cite{moukaddem2022comparison} & DVCKD \cite{dhaun2014optical, farrah2023choroidal} \\
 \midrule
Cohort demographics &  &  &  \\
\cmidrule(l){1-1}
Participants (Eyes) & 60 (120) & 21 (33) & 22 (22) \\
Right eyes (\%) & 60 (50) & 15 (45.5) & 22 (100) \\
Age (SD) & 34.7 (5.2) & 23.9 (4.2) & 21.3 (2.2) \\
Sex, F (\%) & 60 (100) & 9 (43) & 10 (45.5) \\
Ethnicity & 52 White, 6 Asian, 2 Mixed & 12 White, 5 Asian, 2 Black, 2 Middle Eastern & Unknown \\
Refractive status\textsuperscript{\textdagger} & 3 hyperopes, 31 emmetropes, 26 myopes & 9 hyperopes, 8 emmetropes, 3 hyperopes* & Axial length 24.1 $\pm$ 1.3 mm --- mild myopes. \\
Study purpose & Normative / growth restricted / pre-eclamptic & Diurnal variation & Diurnal variation \\
Control/Case & 45/11/4 & 21/0 & 22/0 \\
 &  &  &  \\
Image characteristics &  &  &  \\
\cmidrule(l){1-1}
Device & Spectralis (Heidelberg) & DRI Triton Plus (Topcon) & Spectralis (Heidelberg) \\
OCT Type & Spectral-domain & Swept-source & Spectral-domain \\
Scan Pattern & Macular volume & Macular radial & Peripapillary \\
Mode & HRA+OCT & CFP+OCT & HRA+OCT \\
Time of day (Interval) & All in afternoon (1 minute) & 13 morning, 12 afternoon, 8 evening (5 minutes) & Each at 9am, 12:30pm, 4pm ($\pm$ 40 minutes) \\
B-scans per eye & 31 (EDI) / 61 (non-EDI) & 12 & 1 \\
ART & 50 (EDI) / 12 (non-EDI) & NA & 100 \\
SNR & 35.61 & \textgreater 88 & \textgreater{}25 \\
B-scan image resolution & 496 $\times$ 768 & 992 $\times$ 1024 & 768 $\times$ 1536 \\
\bottomrule
\end{tabular}}}}
\caption{Population demographics and image characteristics of the three samples used to assess OCTolyzer's reproducibility for choroid analysis. \textsuperscript{\textdagger}: Myopic/hyperopic status defined as < -1 / > 1 dioptres. Dioptre measurements were taken from OCT scan metadata (i-Test sample), spherical equivalents (GCU Topcon sample), or axial length (DVCKD sample). *: 1 participant had a missing spherical equivalent measurement.}
\label{tab:rep_table} 
\end{table*}

For i-Test \cite{dhaun2014optical}, data were acquired using the Heidelberg Spectralis SD-OCT Standard and FLEX modules (Heidelberg Engineering, Heidelberg, Germany), collecting two unregistered macula-centred volume scans per eye. Scans covered a 30 $\times$ 20 degree field of view (9 $\times$ 6.6 mm) with enhanced depth imaging (EDI) toggled on and off. EDI volumes comprised 31 B-scans spaced 240 $\mu$m apart with automatic real time (ART) B-scan averaging of 50, while non-EDI volumes included 61 B-scans spaced 120 $\mu$m apart with ART averaging of 12. B-scans had an image resolution of 496 $\times$ 768 (pixel height $\times$ width), with an average signal-to-noise (SNR) score of 35.58.

The GCU Topcon study \cite{moukaddem2022comparison} used the swept-source OCT (SS-OCT) Topcon DRI Triton Plus swept source OCT device (Topcon, Tokyo, Japan) to assess choroidal diurnal variation, primarily recruiting hyperopes. Fovea-centred, 12-line radial OCT B-scans were captured, starting horizontally and rotating in 30 degree intervals. Repeated OCT scans were taken within 5 minutes, with B-scans having an image resolution of 992 $\times$ 1024 pixels and covering 9 mm laterally. Scans with an average SNR below 88 were excluded. Repeated data were available for 12 participants in both eyes and 9 participants in one eye.

In the DVCKD study \cite{dhaun2014optical}, participants were recruited to assess diurnal retinochoroidal changes in relation to chronic kidney disease. SD-OCT peripapillary scans of the right eye were collected using the Heidelberg Spectralis Standard Module with EDI mode on with an ART of 100. Peripapillary B-scans are circular scans with an image resolution of 768 $\times$ 1536 pixels, and centred on the optic disc. B-scans were acquired at three time points: morning (09:12 $\pm$ 12 min), early afternoon (12:36 $\pm$ 7 min), and early evening (16:08 $\pm$ 8 min).

\subsection{OCTolyzer's segmentation module}

\subsubsection{OCT Segmentation}

OCTolyzer does not have a standalone algorithm for retinal layer segmentation, but supports extraction of the segmented layers from the input file metadata, as it's common for OCT manufacturers to have their own built-in segmentation tool for the retinal layers. Whether all retinal layers have been quality checked is at the discretion of the end-user. This is typically performed on the manufacturers' proprietary software, such as the Heidelberg Eye Explorer (HEYEX) viewer (Heidelberg Engineering, Heidelberg, Germany) \cite{manual2022heidelberg}. 

Choroid segmentation for macula-centred OCT B-scans is with Choroidalyzer \cite{engelmann2023choroidalyzer}. Choroidalyzer is a deep learning-based tool which automatically segments the choroidal region and vasculature, and also detects the fovea on fovea-centred OCT B-scans. 

Choroid segmentation for disc-centred, peripapillary B-scans is with DeepGPET \cite{burke2023open}. DeepGPET is a deep learning-based tool which automatically segments the choroidal region, and is robust to segmenting peripapillary choroids of which it was not trained on (Supplementary \cref{suppfig:deepgpet_peripapillary}).

\subsubsection{SLO Segmentation}

There are three segmentation models for OCTolyzer's SLO segmentation module, one for binary vessel detection, another for fovea detection and a final one for segmentation of the en face retinal vessels into arteries and veins, and detection of the optic disc (artery-vein-optic disc detection) \cite{burke2024sloctolyzer}. 

\subsection{OCTolyzer's measurement module}
OCTolyzer's OCT analysis suite supports feature extraction of retinochoroidal spatial measurements on OCT scans, and the SLO analysis suite supports feature extraction of en face retinal vessels in SLO images \cite{burke2024sloctolyzer}.

\begin{figure*}[t]
    \centering
    \includegraphics[width=\textwidth]{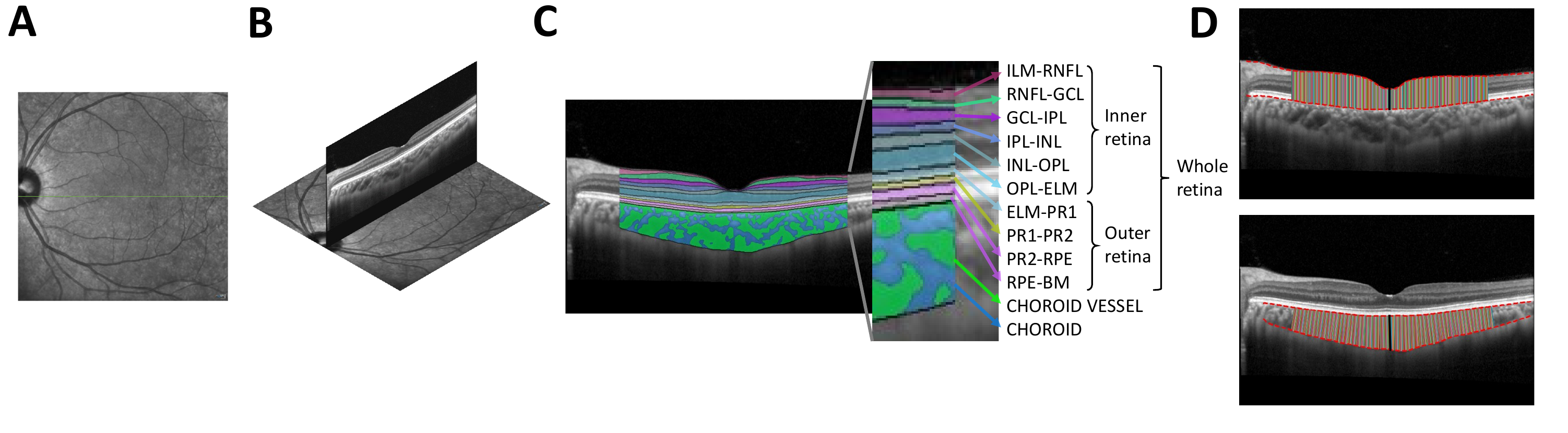}
    \caption{Diagram of single macular OCT B-scan measurements. (A) Localiser SLO showing the OCT acquisition line (green). (B) Horizontal-line OCT B-scan overlaid on the localiser. (C) OCT B-scan with retinal and choroid segmentations labelled (retinal layer definitions in Supplementary \cref{supptab:layers_computed}). (D) Thickness measurements drawn per A-scan for retina (top) and perpendicular to upper boundary for choroid (bottom). The solid black line indicates subfoveal thickness.}
    \label{fig:thickness_diagram}
\end{figure*}

\begin{figure*}[t]
    \centering
    \includegraphics[width=\textwidth]{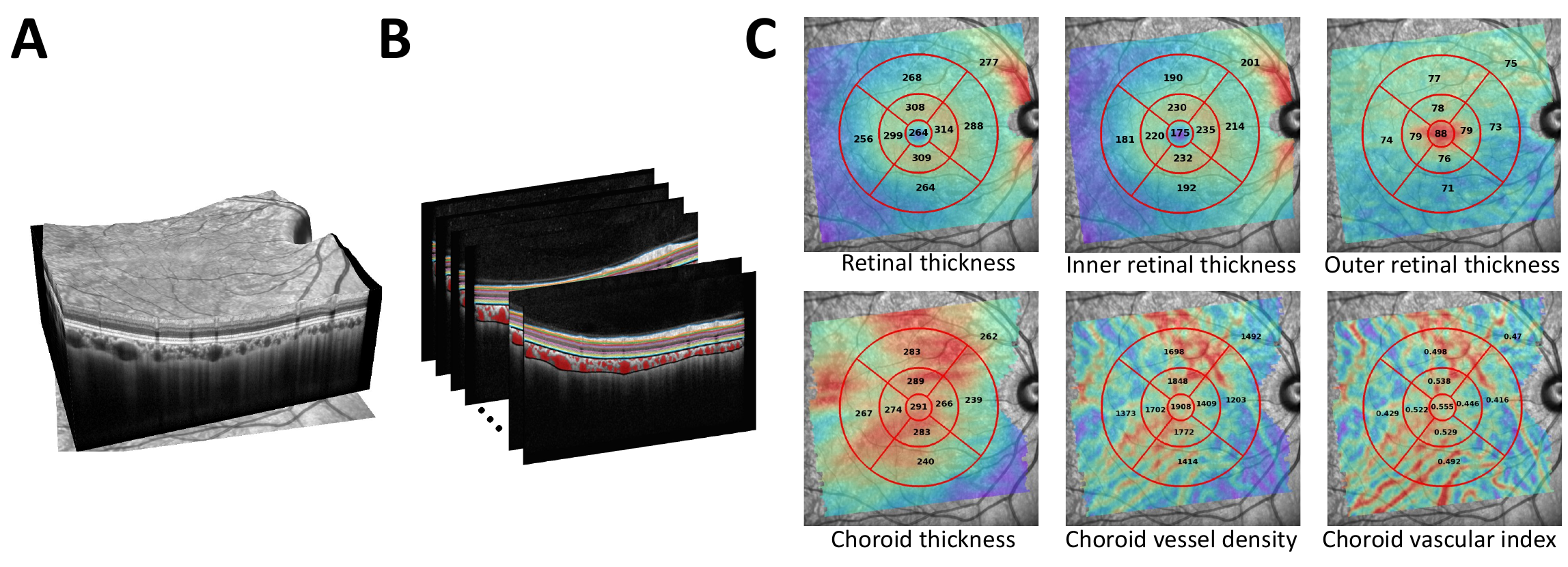}
    \caption{Measurement process for a macular OCT volume. (A) 3D visualisation of an OCT volume scan. (B) Sequential B-scans with retinal and choroid segmentations. (C) Thickness maps for the inner, outer, and whole retina (top), and choroid thickness, vessel density, and CVI (bottom), with average ETDRS grid measurements overlaid. Inner retinal layer thickness maps can also be computed.}
    \label{fig:volume_diagram}
\end{figure*}

\begin{figure*}[t]
    \centering
    \includegraphics[width=\textwidth]{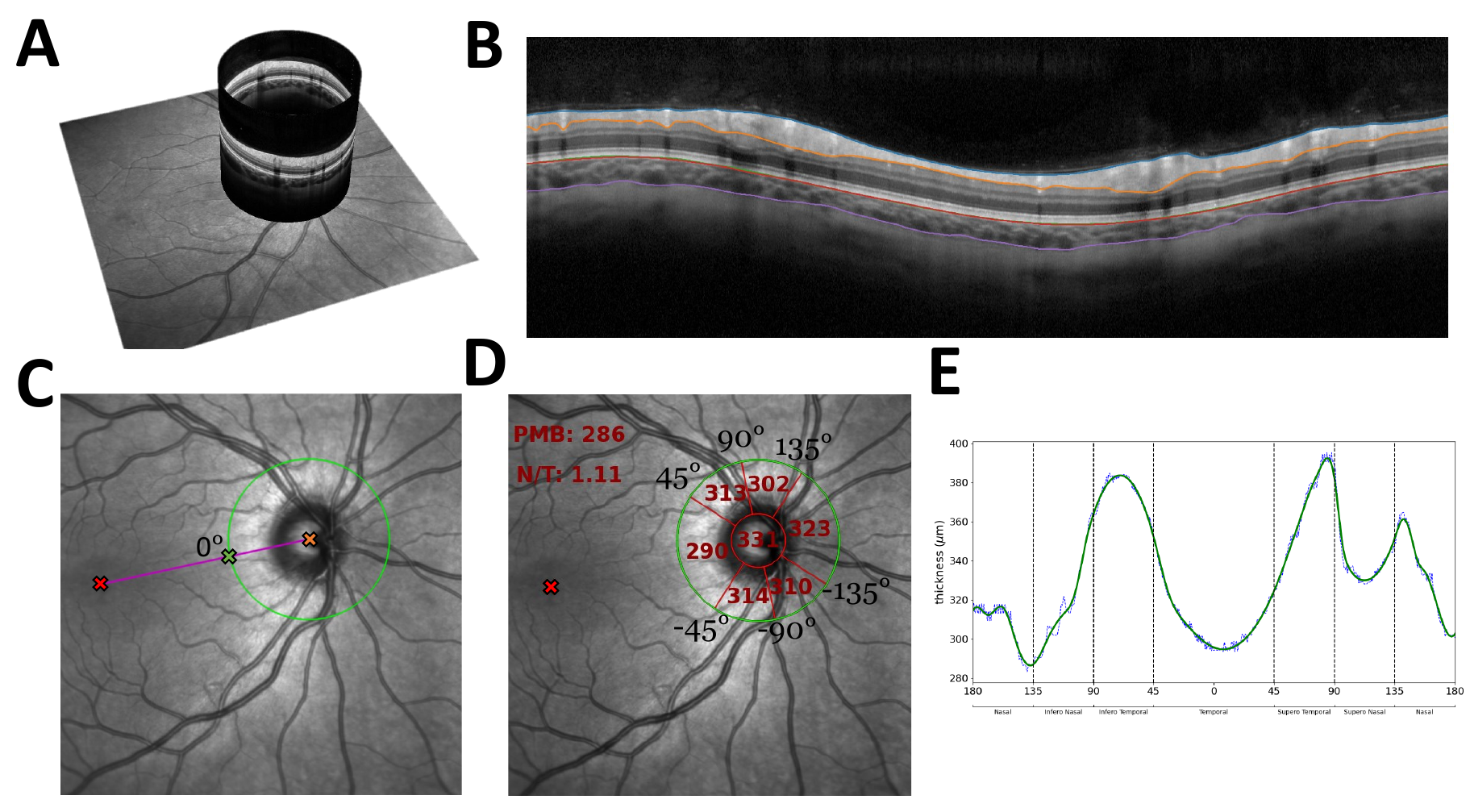}
    \caption{Measurement process for an OCT peripapillary B-scan. (A) Circular OCT B-scan overlaid on the localiser SLO. (B) B-scan with retinal and choroid segmentations. (C) Detection of the temporal sub-field centre. (D) Division into 6 peripapillary sub-fields with retinal measurements overlaid. (E) Peripapillary retinal thickness profile with sub-field thresholds superimposed.}
    \label{fig:peripapillary_diagram}
\end{figure*}

\begin{figure*}[t]
    \centering
    \includegraphics[width=\textwidth]{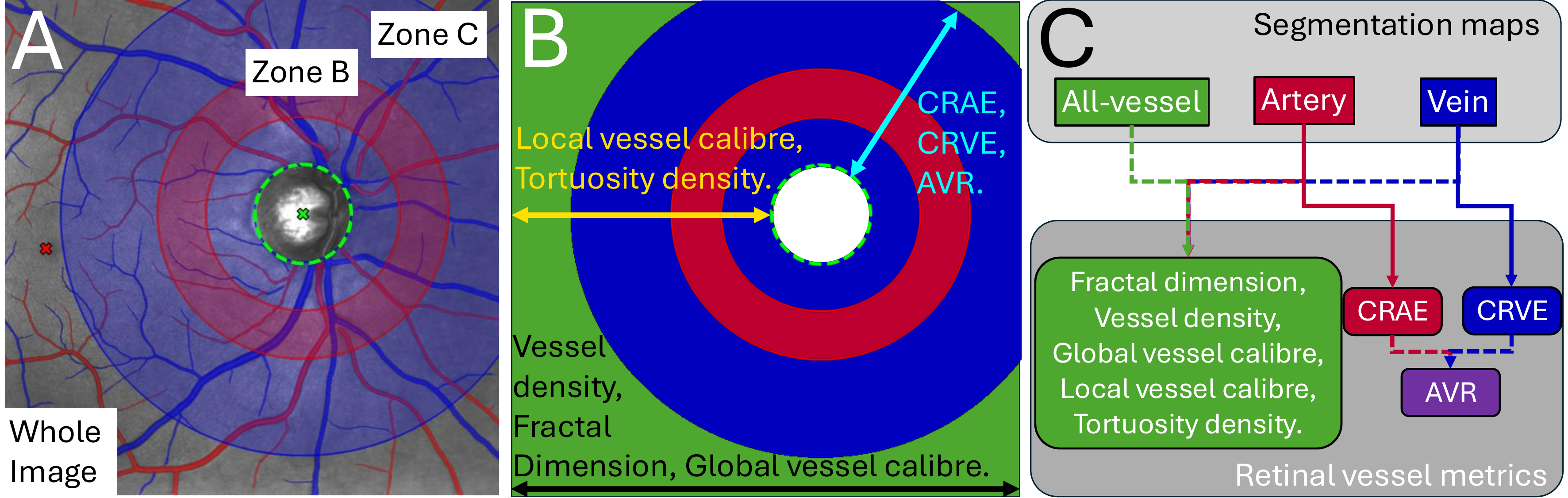}
    \caption{Feature extraction for the localiser SLO image. (A) Disc-centred localiser SLO with segmentations and regions of interest overlaid. (B) Region of interest masks: whole image (green), zone C (blue), and zone B (red), with vessel features indicated by colour-coded arrows extending to the regions which they are measured in. (C) Flowchart of vessel metrics by segmentation map. Figure reproduced and edited with permission from Burke et al. \cite{burke2024sloctolyzer}.}
    \label{fig:sloctolyzer_metrics}
\end{figure*}

\subsubsection{Macular B-scan}

For macular OCT B-scans, Choroidalyzer detects the fovea pixel coordinate for fovea-centred scans. For a user-defined, fovea-centred region of interest (RoI), OCTolyzer measures subfoveal and average thickness, area for all layers, choroidal vessel area, and choroidal vascular index (CVI). Retinal measurements require available segmentations in the file metadata; otherwise, they are excluded.

The choroid is defined as the space posterior to hyper-reflective Bruch's complex and anterior to the sclera. Choroidal measurements account for choroidal curvature by defining RoIs perpendicular to the upper boundary, while retinal measurements define RoIs per A-scan, or vertically. Thus, choroid thickness is measured as the perpendicular distance between upper and lower boundaries, while retinal thickness is measured per A-scan as a vertical micron distance between upper and lower boundaries. Area is calculated as the pixel count within the RoI converted to mm$^2$, and choroidal vascular index as the ratio of choroid vessel area to total choroid area. \cref{fig:thickness_diagram} illustrates an OCT B-scan with segmented layers (C) and overlaid thickness lines for retina and choroid (D).

\subsubsection{Macular volume scan}

A macula-centred OCT volume comprises sequential, parallel B-scans. Choroidalyzer segments the choroidal layer and vessels across all B-scans. The foveal pit is assumed to be present in one or more B-scans and the B-scan and pixel coordinate with the highest probability outputted from Choroidalyzer are selected as the central B-scan and fovea coordinate. The fovea on the localiser SLO is then identified by cross-referencing the fovea coordinate on the relevant B-scan. Retinal measurements require segmentation data in the metadata; otherwise, they are excluded.

From an OCT volume scan, segmentation-derived spatial measurements across the macula visualised as thickness maps are used for feature extraction. Across all B-scans, thicknesses for each layer of interest are measured at every A-scan for the retina and perpendicularly for choroid, and aligned using the detected OCT fovea's lateral position. This creates a coarse two-dimensional map of thickness values. To register the map onto the SLO, these thickness values are interpolated, smoothed, padded, and rotated to align with the localiser SLO. Bi-linear interpolation and Gaussian blur are applied to refine the coarse map and match the SLO image resolution, respectively. Padding centres the map on the detected fovea on the SLO, while rotation aligns it with the OCT volume's acquisition region. Rotation angles are derived from the B-scan location on the SLO. Supplementary \cref{suppfig:thickness_map_diagram} illustrates the pipeline for generating a choroid thickness map.

Using the detected fovea on the SLO as the centre, three concentric circles (1 mm, 3 mm, and 6 mm diameter) are overlaid on the SLO, with the 3 mm and 6 mm circles divided into four quadrants. Quadrants are rotated based on the OCT volume's acquisition angle. Average thickness (in microns) and interpolated volumes (in mm$^3$) are calculated for these 9 sub-regions, following the Early Treatment Diabetic Retinopathy Study (ETDRS) grid \cite{early1991early}, along with a global average for the whole 6 mm diameter RoI. Additionally, choroid vessel density (micron$^2$) and choroid vascular index maps are computed for each sub-field of the grid. Any missing values in the sub-fields are approximated using nearest-neighbour interpolation.

\cref{fig:volume_diagram} outlines the OCT volume processing pipeline, from raw data to measurements.

\subsubsection{Peripapillary B-scan}

For each layer of interest, an array of thickness values are measured as micron distances per A-scan. Choroid thickness is not measured perpendicularly, as its sinuous shape on the B-scan reflects the circular acquisition pattern rather than anatomical curvature. If segmentations do not span the image laterally, the thickness array is linearly interpolated to maintain continuity due to the circular scan pattern.

To define the peripapillary grid, the temporal sub-field centre (0 degrees) is the point along the B-scan's acquisition line co-linear with the optic disc and fovea centres. If the localiser SLO is unavailable, the lateral B-scan centre is used, though this may offset the grid producing maligned measurements. Average thickness per sub-field in the grid are measured by dividing the thickness array into temporal (-45 -- 45 degrees), supero-temporal (45 -- 90 degrees), supero-nasal (90 -- 135 degrees), nasal (135 -- -135 degrees), infero-nasal (-135 -- -90 degrees) and infero-temporal (-90 -- -45 degrees) sub-fields. Metrics include average thickness per sub-field, nasal-to-temporal (N/T) ratio, papillomacular bundle thickness (-30 to 30), and global average. Thickness profiles are drawn which overlay the raw and smoothed thickness arrays with peripapillary grid superimposed. \cref{fig:peripapillary_diagram} illustrates this process.

\subsubsection{localiser SLO}

The features extracted from the en face retinal vessels segmented from the localiser SLO image have been described previously \cite{burke2024sloctolyzer}, and the different RoIs which they are measured in are described in \cref{fig:sloctolyzer_metrics}. In brief, we have adapted the code provided by Automorph \cite{zhou2022automorph} to measure vessel complexity, density, tortuosity and calibre of the arteries, veins and all-vessels from the localiser SLO. We measure fractal dimension (using the Minkowski-Bouligand dimension \cite{falconer2004fractal}), vessel density (ratio of vessel pixels to image resolution), global vessel calibre (ratio of vessel pixels to skeletonised vessel pixels) and local vessel calibre (vessel calibre computed and averaged across individual vessel segments). We also measure central retinal artery and vein equivalents (CRAE, CRVE), computed using the Knudston approach \cite{knudtson2003revised}, as well as tortuosity density \cite{grisan2003novel}. For both macula- and disc-centred SLO images, all aforementioned measurements are computed across the whole image. For disc-centred SLO images, CRAE/CRVE, local vessel calibre and tortuosity measurements are computed for circular RoIs centred on the optic disc and defined using its diameter (D), zone B (0.5D -- 1D) and C (0D -- 2D) \cite{cameron2016modulation}.

\subsection{OCTolyzer's pipeline}

\subsubsection{Input}
OCTolyzer fully supports \verb|.vol| file formats (from Heidelberg Engineering imaging devices) which uses the EyePy python package for file reading \cite{morelle2023eyepy}. There is currently no support for \verb|.e2e| files as current python-based file readers \cite{morelle2023eyepy, graham2024octconverter} are unable to locate the necessary pixel length scales for converting from pixel space into physical space. We do not support other proprietary file formats like \verb|.fda|, \verb|.fds|, \verb|.img| and \verb|.oct|, due to the propensity for storing ophthalmic image and metadata in vendor-neutral formats like DICOM (\verb|.dcm|). \verb|.dcm| file formats currently have limited support but we are working on this presently. See Supplementary \cref{suppfig:octolyzer_input} for information on OCTolyzer's expected input, setup and interface.

\subsubsection{OCT analysis suite}
OCT B-scans are brightened using gamma-level contrast enhancement to set their mean pixel intensity to approximately 0.2, after normalising in [0, 1]. This is especially useful for scans with large choroids or B-scans captured without EDI mode on. Choroidal vessels often experience shadowing when the OCT beam penetrates through superficial retinal vessels sitting perpendicular to the incident laser light, darkening deeper structures \cite{schuman2024optical}. A multiplicative compensation factor is computed and applied which brightens corrupted A-scans, calculated as the ratio of each A-scan's (axially) averaged signal to a laterally smoothed moving average \cite{mao2019deep}.

OCTolyzer's choroid segmentation uses Choroidalyzer \cite{engelmann2023choroidalyzer} for macula-centred B-scans and DeepGPET \cite{burke2023open} for peripapillary B-scans. Peripapillary B-scans are padded laterally with 240 pixels to facilitate end-to-end segmentation. Binary segmentation masks are created using thresholds of 0.5 for Choroidalyzer and 0.25 for DeepGPET, with the latter adjusted for peripapillary B-scans to ensure these are segmented laterally end-to-end. OCTolyzer uses the raw probability map for choroid vessel segmentation outputted by Choroidalyzer. This is to handle poorly defined vessel walls through uncertainty, given the choroid's dense vascular space is often imaged obliquely. Fovea detection in macular B-scans uses 21-wide and 51-long triangular filters to extract the exact pixel coordinate from Choroidalyzer's probability map \cite{engelmann2023choroidalyzer}. Finally, OCTolyzer will extract retinal layer segmentation, if they exist in the metadata. 

OCTolyzer's measurement module will use all layer segmentations to compute spatial measurements across the macula or around the disc. Relevant visualisations of B-scans with segmentations overlaid and thickness maps/profiles overlaid onto the localiser SLO are generated and saved out for downstream analysis and quality assessment --- optionally, to prevent problems of memory consumption during large-scale batch processing. Key features are extracted from the layer segmentations, including thickness/area/volume around the macula/disc for both retina and choroid. 

Any issues during processing, such as missing values, are logged to the end-user. For peripapillary OCT, to assess the quality of scan acquisition, we defined an overlap index to assess the OCT scan centring relative to the detected optic disc, measured as a percentage of the optic disc diameter. Scans exceeding 15\% overlap are suggested as off-centre and logged to the end-user. Supplementary \cref{suppfig:overlap_index_peri} shows examples of scans within and beyond this threshold. 

\subsubsection{SLO analysis suite}
This segmentation module resizes the SLO image to a common image resolution of 768 $\times$ 768 for segmentation, with output maps resized back to the original resolution. Segmentation masks are binarised at a 0.5 threshold, and vessel maps undergo morphological operations to enhance connectivity and remove small false positives. For macula-centred scans, the OCT-derived fovea coordinate is cross-referenced onto the SLO and used as the fovea coordinate. For disc-centred images, the optic disc is modelled as an ellipse, and its diameter is the average of the major and minor axes. The measurement module calculates complexity, tortuosity, and calibre for all vessels, arteries, and veins across the entire image and zones B and C. Optionally, segmentation masks and composite overlays of the segmented retinal vessels, fovea and optic disc are superimposed onto the localiser SLO and saved out. OCTolyzer also allows manual segmentation correction via ITK-Snap \cite{py06nimg}, with features recomputed after re-running OCTolyzer with the manual annotations.

\subsubsection{Output}

Alongside segmentation masks and key visualisations, a process log and extracted features are saved out. Additionally, key metadata extracted from the input file and during processing are stored, such as the average SNR of the B-scans, crucial pixel length scales, coordinates of the fovea on the B-scan and localiser SLO, the centre of the optic disc and its estimated radius. Supplementary \cref{supptab:metadata_keys} presents a full list of the metadata outputted by OCTolyzer after processing a \verb|.vol| file from a Heidelberg Engineering device. See Supplementary \cref{suppfig:octolyzer_output} for information on OCTolyzer's outputs.

\subsection{Statistical analysis}
We assess the reproducibility of OCTolyzer's choroid segmentation models on downstream clinical measurements. Choroidalyzer was used for macular OCT data in the i-Test and GCU Topcon samples, while DeepGPET was applied to peripapillary OCT data in the DVCKD sample.

In the i-Test sample, metrics average choroid thickness and CVI were extracted along all 9 sub-fields of the ETDRS grid, comprising 1080 comparisons per metric (9 features per 120 eyes). For the DVCKD sample, choroid thickness along all 7 sub-fields of the peripapillary grid were extracted for every time point and were compared across consecutive time points (morning -- afternoon, afternoon -- evening), comprising 308 comparisons per metric (2 sets of comparisons for 7 features per 22 eyes). In the GCU Topcon sample, there were several repeated instances per eye. To prevent any sampling bias and remain objective in our reproducibility analysis, we selected one repeated pair per eye and metrics subfoveal choroid thickness (SFCT), choroid area and CVI were extracted for every B-scan within a fovea-centred, 6 mm RoI, comprising 396 comparisons per metric (1 feature per 12 B-scans across 33 eyes).

Reproducibility was measured at the population level using mean, standard deviation (SD), mean absolute error (MAE), and correlation metrics (Pearson, Spearman, ICC(3,1)), supported by correlation and Bland-Altman \cite{bland1986statistical} plots. At the eye level, reproducibility was assessed using measurement noise estimate $\lambda$ \cite{engelmann2024applicability}, which is a way to compare measurement variability (noise) within a single eye (within-eye variability) to variability between different eyes (between-eye variability).

For each metric, $\lambda$ is the ratio between the SD of within-eye measurements and the SD of between-eye measurements. To calculate this for every eye, we first average the features for each member of every eye's repetition pair. We then measure the SD between each repeated pair (within-eye variability) and the SD across all eyes (between-eye variability). For example, the within-eye and between-eye variability for OCT volume scans is computed by first averaging the 9 features in the ETDRS grid into a single representative feature for each member of every eye's repetition pair. We then measure the SD for the repeated pair (within-eye variability) and across all eyes (between-eye variability) and take their ratio. A similar approach is done for all 6 peripapillary sub-fields and 12 B-scans for the DVCKD and GCU Topcon samples, respectively. $\lambda$ is measured for every eye and is presented as a percentage for convenience, with 0 as the optimal value. 

Finally, OCTolyzer’s execution time was measured for its full analysis pipeline, with and without the SLO suite, using 100 trials per OCT data type, reporting mean and SD times in seconds. Timed experiments were run on a Windows laptop with a 4-year-old Intel Core i5 (8th generation) CPU and 16 Gb of RAM. For brevity, we will refer to this as the ``laptop CPU'' in the rest of the text. 

\begin{table*}[t]
\centering
\begin{adjustbox}{max width=\textwidth}
{\small\scalebox{0.99}{\begin{tabular}{rllllll}
\toprule
\multicolumn{1}{l}{Method} & Metric [unit] & Mean (SD) & MAE & Pearson & Spearman & ICC(3,1) \\
\midrule
\multicolumn{1}{c}{Choroidalyzer} &  &  &  &  &  &  \\
\cmidrule(l){1-2}
\multirow{1}{*}{i-Test} & CT [$\mu$m] & 274.1 (91.8) & 6.7 & 0.99 & 0.99 & 0.99 \\
 & CVI & 0.51 (0.03) & 0.027 & 0.97 & 0.97 & 0.98 \\
 \cmidrule(l){2-7}
\multirow{1}{*}{GCU Topcon} & SFCT {[}$\mu$m{]} & 392.6 (110.9) & 11.6 & 0.99 & 0.99 & 0.99 \\
 & CA {[}mm$^2${]} & 1.62 (0.46) & 0.051 & 0.98 & 0.99 & 0.99 \\
 & CVI & 0.53 (0.03) & 0.013 & 0.91 & 0.91 & 0.95 \\
 \midrule
\multicolumn{1}{c}{DeepGPET} &  &  &  &  &  &  \\
\cmidrule(l){1-2}
\multirow{1}{*}{DVCKD} & CT {[}$\mu$m{]} & 169.6 (54.1) & 5.0 & 0.99 & 0.99 & 0.99 \\
\bottomrule
\end{tabular}}}
\end{adjustbox}
\caption{Reproducibility performance for OCTolyzer's OCT analysis suite. All Pearson and Spearman correlations were statistically significant with P-values $P < 0.0001$.}
\label{tab:octolyzer_repr_tab} 
\end{table*}

\begin{figure*}[t]
    \centering
    \includegraphics[width=\textwidth]{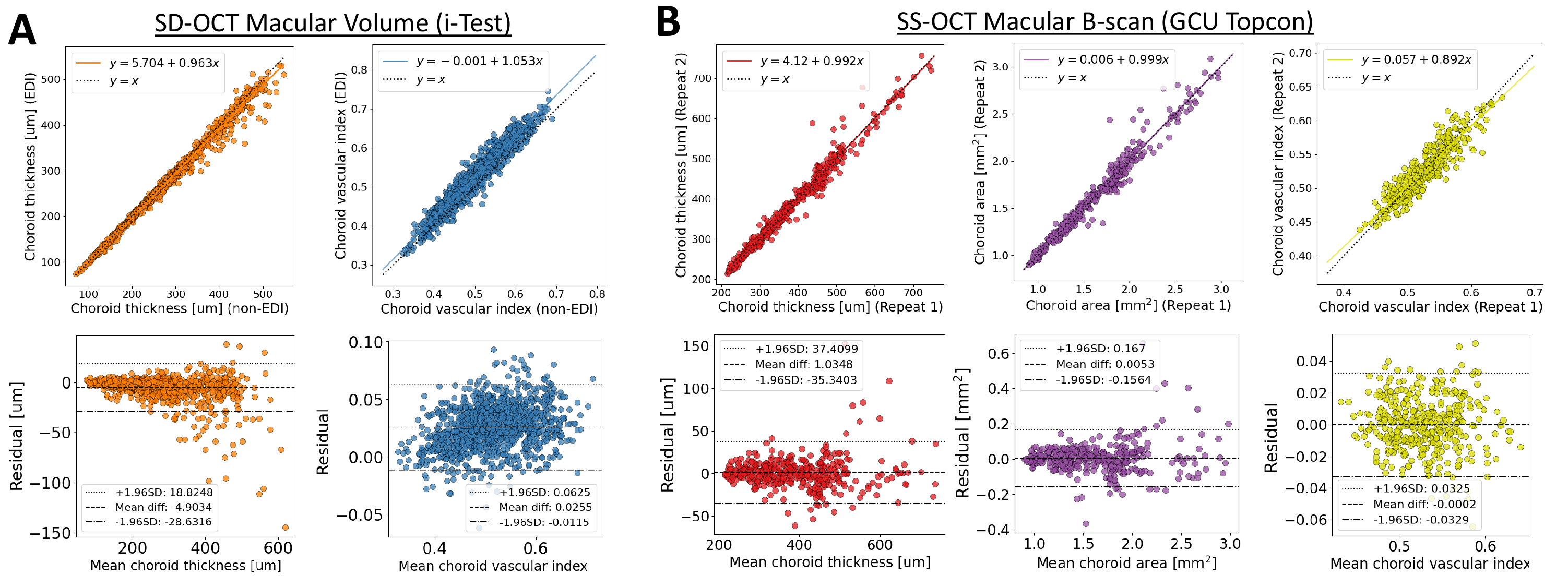}
    \caption{Correlation and Bland-Altman plots for assessing the reproducibility of OCTolyzer's OCT analysis suite for macular OCT data. (A) Macular volume scan pairs from the i-Test sample. (B) Macular B-scan pairs from the GCU Topcon sample.}
    \label{fig:octolyzer_pop_repr_macular}
\end{figure*}

\begin{figure*}[t]
    \centering
    \includegraphics[width=\textwidth]{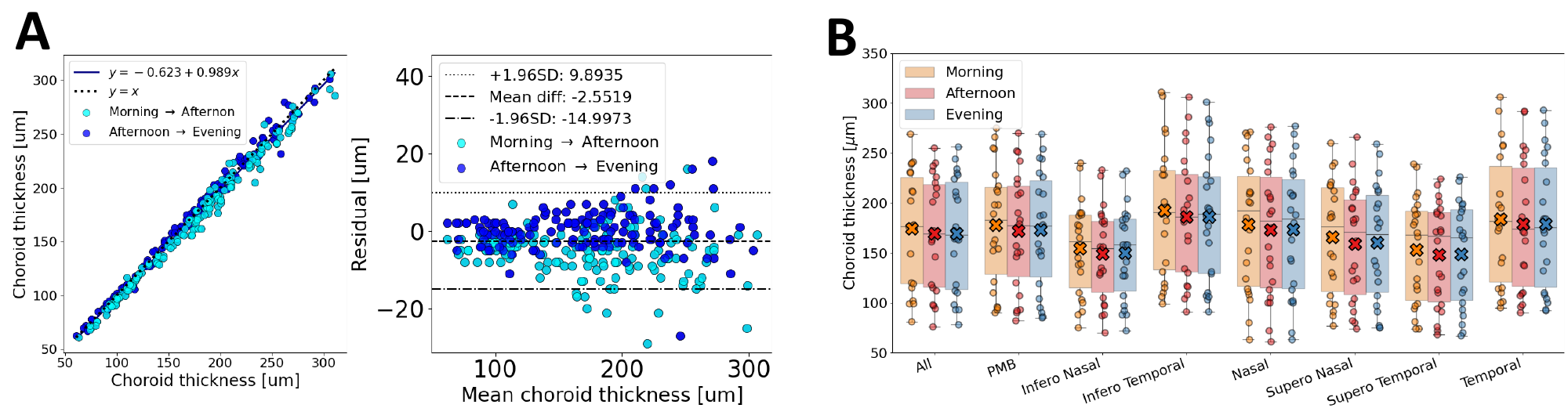}
    \caption{(A) Correlation and Bland-Altman plot for all chronologically paired (morning -- afternoon and afternoon -- evening) choroid thickness measurements in all peripapillary sub-fields. (B) Longitudinal evolution of average choroidal thickness in each peripapillary sub-field shown as box-plots, with mean values overlaid as crosses.}
    \label{fig:octolyzer_pop_repr_perip}
\end{figure*}

\begin{figure*}[t]
    \centering
    \includegraphics[width=\textwidth]{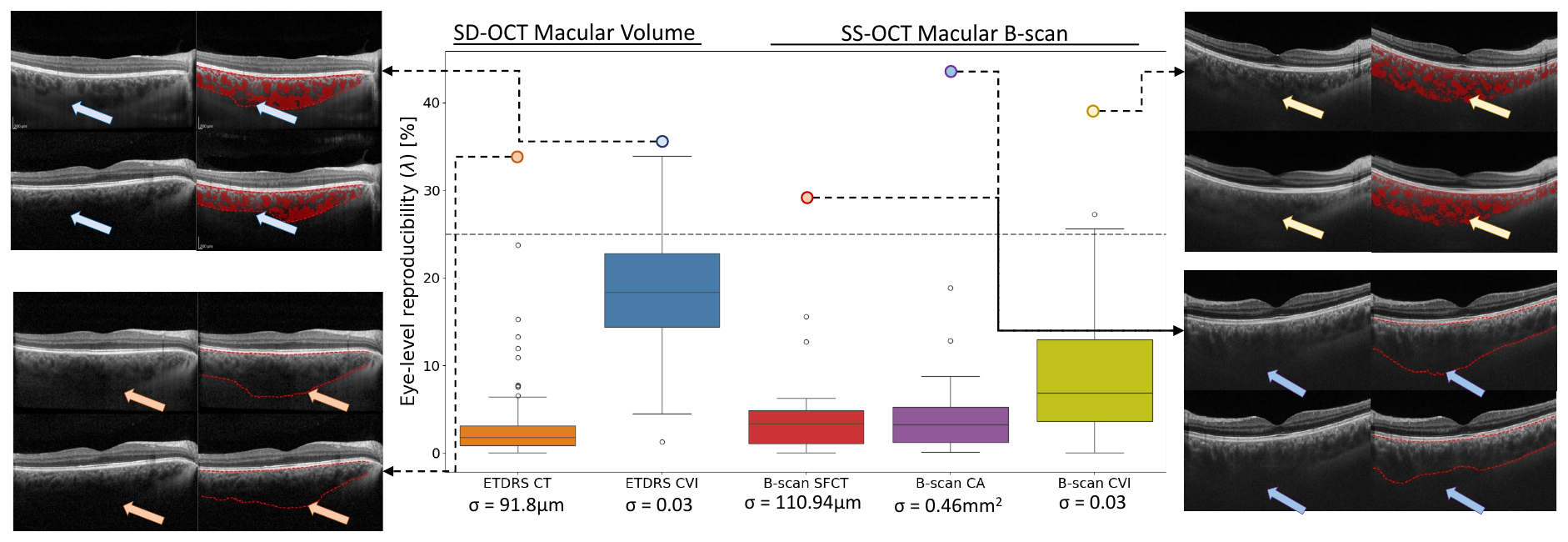}
    \caption{Reproducibility of OCTolyzer's OCT analysis suite for macular OCT data at the eye level. Major outlier B-scans are shown for all metrics, with segmentations overlaid in red and arrows indicating the source of error. The between-eye SDs are shown below each box-plot.}
    \label{fig:octolyzer_macular_repr}
\end{figure*}

\begin{figure*}[t]
    \centering
    \includegraphics[width=\textwidth]{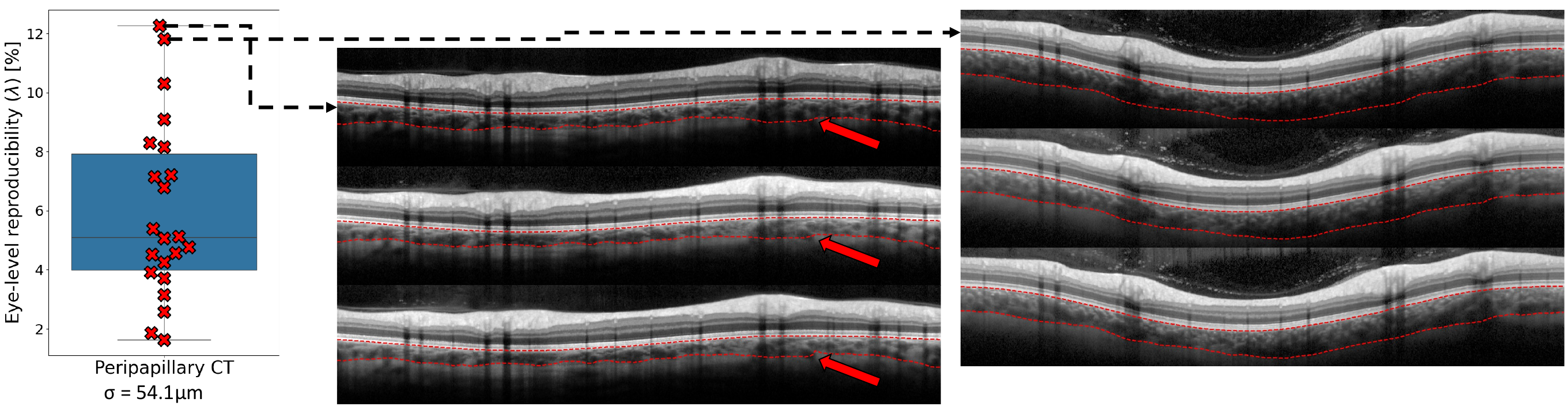}
    \caption{Reproducibility of OCTolyzer's OCT analysis suite for OCT peripapillary CT data at the eye level. Red crosses are the individual values for $\lambda$ across all 22 eyes. Major outlier B-scans are shown to the right for all time points, with segmentations overlaid in dashed red and red arrows showing any obvious sources of error.}
    \label{fig:octolyzer_peri_repr}
\end{figure*}

\section{Results}

\subsection{Reproducibility} 

\subsubsection{Population level}
\cref{tab:octolyzer_repr_tab} summarises the reproducibility performance of OCTolyzer's OCT analysis suite across three supported OCT data types. Correlations for choroid segmentation metrics were high in all cohorts (e.g., Pearson/Spearman for ETDRS choroid thickness (CT): 0.99/0.99, B-scan SFCT: 0.99/0.99, B-scan choroid area (CA): 0.98/0.99, peripapillary CT: 0.99/0.99). Reproducibility for CVI was slightly lower but remained strong (ETDRS CVI: 0.97/0.97, B-scan CVI: 0.91/0.91).

In the GCU Topcon sample, subfoveal choroids were thicker (mean $\pm$ SD: 392.6 ± 110.9 $\mu$m) compared to the i-Test sample (274.1 $\pm$ 91.8 $\mu$m) and the DVCKD sample had the thinnest choroids (169.6 $\pm$ 54.1 $\mu$m). Observed differences between cohorts reflect variations in demographics, scanning locations and imaging/measurement methodology. For example, the GCU Topcon cohort targeted hypermetropes during recruitment \cite{moukaddem2022comparison}, resulting in smaller eyes and thicker choroids. Moreover, the i-Test sample represent ETDRS sub-field values which are averages taken over different-sized regions in the macula, allowing for noise in measurement error to be averaged out unlike one-dimensional, point-source SFCT. Additionally, peripapillary choroids are thinner near the optic disc than in the central macula \cite{yang2019factors}.

These cohort differences explain the variation in MAE in CT, with higher values observed in the GCU Topcon sample (11.6 $\mu$m) compared to DVCKD (5.0 $\mu$m) and i-Test (6.7 $\mu$m). Additionally, across macular data, larger choroids corresponded to larger error in regional metrics, as illustrated by the Bland-Altman plots in \cref{fig:octolyzer_pop_repr_macular}. For macular CVI, the MAE for B-scan CVI in the GCU Topcon sample (0.013) was more than half the ETDRS CVI in the i-Test sample (0.027), likely due to increased noise in non-EDI volume scans. In i-Test, the comparison between EDI and non-EDI volumes translated into systematic underestimation of CT and subsequent overestimation of CVI in the i-Test sample, as illustrated in \cref{fig:octolyzer_pop_repr_macular}(A). In contrast, error in CVI for the GCU Topcon sample showed a centred distribution without bias and no discernible pattern (\cref{fig:octolyzer_pop_repr_macular}(B)).

\cref{fig:octolyzer_pop_repr_perip}(A) presents correlation and Bland-Altman plots for peripapillary CT measurements comparing consecutive time points. Residuals comparing afternoon and evening are centred around 0, while residuals comparing morning and afternoon appear almost consistently below the origin (and identity line in the correlation plot). \cref{fig:octolyzer_pop_repr_perip}(B) shows the longitudinal evolution of peripapillary CT across sub-fields with morning measurements consistently higher than afternoon and evening values.

\subsubsection{Eye level}

\cref{fig:octolyzer_macular_repr} and \cref{fig:octolyzer_peri_repr} present the reproducibility of OCTolyzer's OCT analysis suite for macular and peripapillary OCT data at the eye level, respectively. We present box plots to describe the distribution of measurement noise $\lambda$ \cite{engelmann2024applicability}, expressed in terms of the overall population's variability (shown below the box-plots as the value $\sigma$). 

For choroid region metrics, OCTolyzer's eye level measurement noise was very low compared to the populations' natural variability, with the upper quartile of the box-plot distributions for ETDRS CT/B-scan SFCT/B-scan CA sitting below 5\% of each metrics population SD, and below 10\% for peripapillary CT. In the latter case, the eye level variability is confounded with natural diurnal variation of the choroid, and the population SD for the DVCKD sample was about half of the values for the i-Test and GCU Topcon samples. Thus, for choroid region metrics, segmentation error contributes to only 5-10\% of the overall population's variability. We observed higher variability in choroid vessel metrics at the eye level, with B-scan CVI measurement noise primarily sitting below 15\% of the population SD ($\sigma$=0.03) and 25\% for ETDRS CVI ($\sigma$=0.03). Encouragingly, major outliers for macular OCT B-scans (\cref{fig:octolyzer_macular_repr}) are thick choroids with obscure choroid-sclera boundaries and poor quality vessel wall definition due to decaying optical signal.

In \cref{fig:octolyzer_peri_repr}, we show the B-scans of the major outliers for peripapillary CT at each time point, with segmentations overlaid. Any differences (red arrows) are minimal, and the segmentations do not look qualitatively different. These two cases were the major outliers due to the difference in CT observed from morning -- afternoon (a negative differential of 15 and 14 $\mu$m in average CT, respectively) relative to their afternoon -- evening change (a positive differential of 1 and 2 $\mu$m, respectively). 

\subsection{Execution time}

\begin{table*}[t]
\centering
{\small
\scalebox{0.99}{\centerline{\begin{tabular}{clll}
\toprule
\multicolumn{1}{l}{\multirow{2}{*}{}} & \multicolumn{3}{c}{OCT data type}    \\
\cmidrule{2-4}
\multicolumn{1}{l}{}                  & Macular B-scan        & Macular volume         & Peripapillary B-scan   \\
\midrule
OCT image resolution                        & $1 \times 496 \times 768$ & $61 \times 496 \times 768$ & $1 \times 768 \times 1536$ \\
SLO image resolution                        & $768 \times 768$          & $768 \times 768$           & $1536 \times 1536$         \\
\midrule
Choroid segmentation                  & 0.31 (0.02)               & 16.8 (1.60)                & 1.56 (0.23)                \\
OCT analysis suite                    & 1.55 (0.13)               & 85.00 (3.65)               & 2.29 (0.14)                \\
OCT+SLO analysis suite                & 22.40 (0.97)              & 109.00 (3.11)              & 128.00 (6.37) \\
\bottomrule
\end{tabular}}}}
\caption{Average (SD) execution time in seconds for OCTolyzer's pipeline.}
\label{tab:octolyzer_times} 
\end{table*}

\cref{tab:octolyzer_times} describes OCTolyzer's execution time on all supported OCT data types. Using a laptop CPU, OCTolyzer was able to fully process a single line OCT B-scan in 1.55 $\pm$ 0.1 seconds (image resolution $496 \times 768)$. An OCT volume with thickness maps computed and saved out, with features extracted for every retinal and choroidal layer takes 85 $\pm$ 3.7 seconds (image resolution $61 \times 496 \times 768$). Execution time would improve if the choroid was measured per A-scan (64 $\pm$ 2.5 seconds), rather than locally perpendicular to its upper boundary, and significantly so with GPU acceleration.

An OCT peripapillary scan can be fully processed in 2.29 $\pm$ 0.1 seconds (image resolution $768 \times 1536)$, and takes 1.56 $\pm$ 0.2 seconds to segment the choroidal layer. However, to align the peripapillary grid, the localiser SLO must be available and the fovea and optic disc must be segmented, which increases execution time to 30.70 $\pm$ 2.1 seconds. 

Execution time increases when the SLO analysis suite is toggled on (\cref{tab:octolyzer_times}, final row), particularly for disc-centred SLO images (OCT peripapillary scans), because there are three regions of interest measured for each vessel type (all vessels, arteries, veins). 

\section{Discussion}

We introduced and assessed the reproducibility of an open-source and fully automatic toolkit, OCTolyzer, for segmentation and feature extraction in OCT/SLO data. OCTolyzer's OCT analysis suite demonstrated excellent reproducibility, which was consistent with the reproducibility of OCTolyzer's SLO analysis \cite{burke2024sloctolyzer}. OCTolyzer is capable of producing reproducible and clinically meaningful retinochoroidal features of the back of the eye. Advantageously, the reproducibility analysis conducted provides an upper bound on OCTolyzer's measurement variability which can be used to differentiate true biological change from measurement error, a crucial step in the interpretation of ocular measurements in clinical studies \cite{breher2020choroidal}.

There is a growing recognition at the intersection of healthcare and artificial intelligence (AI) of the need for publicly available methods that have undergone rigorous evaluation \cite{ciobanu2024critical}. The distinct lack of such methods contributes to the current gap between academic research and clinical AI applications \cite{weissler2021role}. Indeed, there is a distinct lack of transparency in the field surrounding their reproducibility, validation strategies and publicly available source code. McDermott, et al. \cite{mcdermott2021reproducibility} surveyed these themes for 511 machine learning studies presented at relevant conferences between 2017 and 2019, of which 211 were in the context of healthcare. Of these 211, they found that only 44\% reported their measurement variability and 21\% made their code and model available. We present OCTolyzer not only as an open-source toolkit which is easy-to-use and accessible to all in the research community, but a toolkit which has undergone significant reproducibility testing, alongside rigorous validation for each of its segmentation models \cite{burke2024sloctolyzer, engelmann2023choroidalyzer, burke2023open}.

For all supported data types at the population level, OCTolyzer had very high correlations across repeated measurements (Pearson, Spearman and ICC(3,1) correlations for regional metrics were all greater than 0.98, and 0.91 for vessel metrics). Increasing error with thicker choroids was expected due to optical signal degradation with axial depth which impacted macula-centred samples primarily --- the GCU Topcon sample reported the largest choroid with an SFCT of 756 $\mu$m (approximately 3.25 SDs away from the mean). 

Importantly, regional metric MAEs were below any change expected due to diurnal variation (approximately 30 $\mu$m in amplitude) \cite{chakraborty2011diurnal, tan2012diurnal, usui2012circadian, kinoshita2017diurnal, singh2019diurnal, ostrin_imi-dynamic_2023} and also below small effect sizes in pathology, such as in myopia progression (20 -- 30 $\mu$m) \cite{breher2019metrological, flores2013relationship}. 

It is also unlikely that the reported CVI MAEs are clinically significant. Breher, et al. \cite{breher2020choroidal} tested the reproducibility of a popular approach to choroid vessel segmentation \cite{sonoda2014choroidal} across different sub-fields of the ETDRS grid and reported a mean difference ranging from 0.039 to 0.051. Additionally, a major review on CVI as a biomarker using another well adopted approach to choroid vessel segmentation \cite{agrawal2016choroidal} in retinal pathology reported changes between healthy and diseased eyes between 0.02 and 0.06 \cite{agrawal2020exploring}.

The systematic bias observed in the macular OCT volume data was due to poor optical signal from each pair's non-EDI volume scan. The under-prediction of choroid thickness in the non-EDI measurements had a direct consequence on ETDRS CVI, as this metric is normalised by the size of the choroid, leading to over-prediction of ETDRS CVI. However, the reported systematic bias of 0.026 for ETDRS CVI is still unlikely to be clinically significant because it sits at the lower bound of previously reported effect sizes in retinal pathology \cite{agrawal2020exploring}.

At the eye level, regional metrics had lower measurement noise (within 5 -- 10\%) than vessel metrics (within 15 -- 25\%), with ETDRS CVI having higher measurement noise than B-scan CVI. The higher measurement noise in ETDRS CVI is potentially due to different scanning parameters (\cref{tab:rep_table}, ART, EDI mode, B-scans per eye) and the unregistered nature of the EDI and non-EDI volume pairs. Nevertheless, while the population variability is notably smaller for CVI measurements, the higher measurement noise could suggest that choroid vessels metrics are less reliable than regional metric, a conclusion drawn also by Breher, et al. \cite{breher2020choroidal}. 

It's very possible that high measurement noise in CVI, relative to regional metrics, is due to the combined error from both vessel and region segmentation, as observed by the major outliers (\cref{fig:octolyzer_macular_repr}, top-left). Therefore, purely vascular metrics like vessel area and vessel volume may be more reliable than CVI and are worth reporting alongside. Fortunately, OCTolyzer supports feature extraction using all aforementioned metrics.

Nonetheless, it's encouraging to observe major outliers representing significant challenges in region and vessel segmentation due to optical signal degradation. In a clinical study, many of the poor quality instances of repeated B-scan pairs would likely be considered for exclusion due to issues with image quality. 

For peripapillary OCT B-scans, analysis of residuals suggested that the major outliers were sourced primarily between morning -- afternoon comparisons, rather than afternoon -- evening comparisons. This corresponds to the trend observed in measuring the same choroids manually in the macula \cite{farrah2023choroidal} (shown in their Supplementary Fig. 8(C)), as well as from previous works studying the natural diurnal variation of the choroid \cite{chakraborty2011diurnal, tan2012diurnal, usui2012circadian, kinoshita2017diurnal, singh2019diurnal, ostrin_imi-dynamic_2023}. These previous studies have reported choroidal fluctuations over the course of day (in the macula and peripapillary regions) with an amplitude of approximately 30 $\mu$m, with the majority reporting higher CT measurements in the morning, compared to the afternoon and evening. Moreover, major outliers at the eye-level showed no qualitative difference in segmentation error.

Thus, given that the largest residuals were those observed between morning and afternoon, and that we know the choroid thins during this time period, we propose that the reproducibility results in the DVCKD sample were confounded by natural diurnal variation of the choroid, and that this was the primary driver of the observed error in the peripapillary data, and not segmentation error.

OCTolyzer's runtime is entirely reasonable for large-scale ophthalmic image analysis, even on a laptop CPU, taking under 2 seconds for a single 496 $\times$ 768 B-scan, just over 2 seconds for a single 496 $\times$ 1536 peripapillary B-scan, and around 85 seconds for a 61 $\times$ 496 $\times$ 768 OCT volume scan. AlzEye is a recently collected, large-scale dataset of retinal images from the Moorefield's Eye Hospital NHS Foundation Trust \cite{wagner2022alzeye}. AlzEye contains 1,567,358 OCT images from 154,830 patients. Assuming the 1,567,358 images were from approximately 25,695 61-stack OCT volume scans, these could be fully processed in approximately 25 days on the same laptop CPU --- an upper bound which could be significantly improved upon with GPU acceleration and stronger compute resources.

Our reproducibility analysis of the OCT suite did have some limitations. Macular OCT volumes in the i-Test sample were not registered between acquisitions, making it difficult to isolate segmentation as the sole source of measurement error. Additionally, in the DVCKD sample, peripapillary choroid measurements were influenced by diurnal variation which confounded the observed measurement error. While this was mitigated in the GCU Topcon sample by using data from a single time point, it was not possible for DVCKD.

Data overlap with Choroidalyzer's training set may have introduced potential bias into the reproducibility analysis. 22\% of the i-Test participants (13/60) and all eyes from the GCU Topcon sample featured in the form of at least one or more B-scans in Choroidalyzer's training set. However, for the i-Test sample this only included EDI OCT volume data, and so the image artefacts introduced by the non-EDI OCT volume data still posed a significant challenge to assess for these 26 eyes. For the GCU Topcon sample, the majority of B-scans from 62\% of participants (13/21) were excluded from Choroidalyzer's dataset because of failure to generate ground-truth segmentation labels during data curation \cite{engelmann2023choroidalyzer}.

OCTolyzer itself has limitations. It was developed for analysing proprietary \verb|.vol| files exported from HEYEX, which typically include the localiser SLO image. Other file types, like \verb|.dcm|, may not contain the SLO image which can harm clinical interpretability (for macular volume data) and accuracy in feature extraction (for peripapillary data). Additionally, OCTolyzer does not perform automatic retinal layer segmentation, assuming instead its availability in the file metadata. Lastly, OCTolyzer's segmentation module was developed using systemic health-related OCT data, not eye pathology. Thus, including deep learning-based retinal layer segmentation and fine-tuning OCTolyzer's segmentation module on pathological data would likely improve OCTolyzer's applicability to the clinical research community.

OCTolyzer also does not automatically assess image or segmentation quality. However, it does provide tools to support end-users. For example, it reports the SNR of OCT data and uses an overlap index to flag off-centre peripapillary scans. Key visualisations optionally saved out allow for real-time quality inspection, and manual annotation of the segmentations from the SLO images is supported (though OCT segmentations cannot be manually edited currently). The process log also provides end-users with key information during processing, such as incomplete segmentations which may lead to missing values. 

Nevertheless, the ability to reject images based on quality or segmentation accuracy is crucial for forming a meaningful dataset of clinical measurements. Current SNR metrics provided by imaging devices do not account for the specific context in which the images will be analysed. Such context-dependent image quality assessment methods is essential for large-scale ophthalmic image analysis and this will be explored in future work. 

Finally, localiser en face images for OCT acquisition are not always an SLO image, but could be a CFP image or other en face modalities. Currently, OCTolyzer supports segmentation and feature extraction only of SLO images. Providing support to process other localiser image types would further enhance OCTolyzer's applicability across different imaging devices and clinical contexts.

\section{Conclusion}
OCTolyzer is the first open-source, fully automatic toolkit that allows high-volume reproducible feature extraction from OCT/SLO data --- an essential ophthalmological modality very commonly used in hospital and by community optometrists. This method greatly improves transparency, speed and standardisation over previous methods. We anticipate the strategic importance of OCTolyzer as an open-source, accurate, and fully-automated image analysis tool in the growing field of oculomics which relates ophthalmic markers to systemic health and disease.
 
OCTolyzer is easy-to-use and can be freely downloaded without author permissions, specialist training or proprietary software. Ultimately, we hope OCTolyzer will facilitate the standardised reporting of ocular measurements, enable large-scale ophthalmic image analysis among the research community, and become a foundational pipeline which the wider research community may continue to extend on and improve, thus addressing the distinct lack of such available pipelines for OCT/SLO data.

\section*{Acknowledgements}
J.B. was supported by the Medical Research Council (grant MR/N013166/1) as part of the Doctoral Training Programme in Precision Medicine at the Usher Institute, University of Edinburgh. This work was partially supported as part of the Medical Research Council Confidence in Concept Scheme (2019), awarded to co-authors I.M. and N.D. Data collection for i-Test was funded by the Wellcome Leap In Utero programme and was not involved in designing, conducting, or submitting this work.

For the purpose of open access, the authors have applied a creative commons attribution (CC BY) licence to any author accepted manuscript version arising. 

\section*{CRediT authorship contribution statement}
Jamie Burke: Conceptualization, Methodology, Software, Validation, Formal analysis, Investigation, Writing – original draft, review \& editing, Visualization. Justin Engelmann: Methodology, Resources. Samuel Gibbon: Resources, Data Curation. Charlene Hamid: Resources, Data Curation. Diana Moukaddem: Data Curation. Dan Pugh: Data Curation. Tariq Farrah: Data Curation, Writing - review \& editing. Niall Strang: Supervision, Writing - review \& editing. Neeraj Dhaun: Supervision. Thomas J MacGillivray: Project administration, Supervision, Writing - review \& editing. Stuart King: Supervision, Writing - review \& editing. Ian J.C. MacCormick: Supervision, Writing - review \& editing.

\section*{Declaration of Interests}
The authors declare no competing financial interests or personal relationships that could have appeared to influence the work reported in this paper.

\section*{Declaration of Generative AI and AI-assisted technologies in the writing process}
The authors declare that no generative AI or AI-assisted technologies were used in the writing of this manuscript.

\bibliographystyle{unsrt}
\bibliography{references}

\onecolumn
 
\appendix
 
\setcounter{figure}{0}
\renewcommand{\thefigure}{S\arabic{figure}}
\setcounter{table}{0}
\renewcommand{\thetable}{S\arabic{table}}
\pagebreak

\section*{Supplementary Material}

\subsection{Simultaneous OCT + SLO capture during acquisition}
\begin{figure}[tbh]
    \centering
    \includegraphics[width=0.8\textwidth]{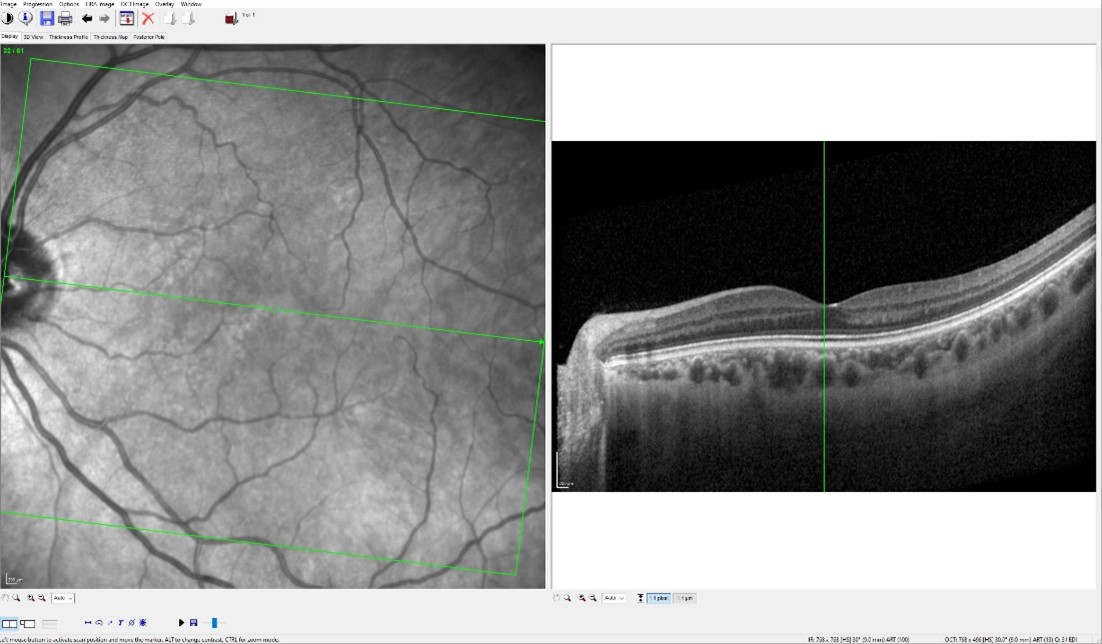}
    \caption{Screenshot from HEYEX viewer of an OCT volume during acquisition. (Left) the SLO image with the acquisition location of the OCT volume overlaid in green, with the green central line indicating the location of the cross-sectional OCT B-scan. (Right) the corresponding fovea-centred OCT B-scan of the OCT volume.}
    \label{suppfig:heyex_demo}
\end{figure}

Supplementary \cref{suppfig:heyex_demo} shows a screenshot from the Heidelberg Eye Explorer (HEYEX) software (version 1.12.1.0) (Heidelberg Engineering, Heidelberg, Germany) of an OCT volume for an individual’s eye. During OCT capture, the SLO image (left) is used to reference the location of the B-scans (right) captured during acquisition. 

\newpage
\subsection{Population tables for OCTolyzer's segmentation models}

\subsubsection{OCT Segmentation}
The following datasets were used for OCT choroid segmentation: OCTANE \cite{dhaun2014optical}, a longitudinal cohort of kidney donors and transplant recipients (47 eyes). Diurnal Variation for Chronic Kidney Disease (DVCKD) \cite{dhaun2014optical}, a cohort of young healthy adults collected to assess diurnal variation of the choroid in Edinburgh (20 eyes). Normative, a collection of OCT data from healthy subjects (author J.B. and a healthy cohort collected for a study related to multiple sclerosis \cite{pearson2022multi}) (60 eyes). i-Test \cite{dhaun2014optical}, a cross-sectional cohort of women undergoing normative, pre-eclamptic or intra-uterine growth restrictive pregnancy (42 eyes). PREVENT Dementia, a cohort of mid-life individuals, half of whom are at genetic risk of developing later life dementia \cite{ritchie2012prevent, ritchie2013prevent} (232 eyes). GCU Topcon, a longitudinal cohort of young, healthy participants with varying degrees of myopia collected for assessing diurnal variation of the choroid in Glasgow (43 eyes). 

For DeepGPET's model construction \cite{burke2023open} only the OCTANE, Normative and a subset of the i-Test data (10 eyes, 5 participants) were used. More details can be found in the original papers describing the methods \cite{burke2023open, engelmann2023choroidalyzer}.

\begin{table}[tbh]
\centering
{\small
\scalebox{0.7}{\centerline{\begin{tabular}{p{2cm}p{2cm}p{2cm}p{2cm}p{2cm}p{2cm}p{2cm}|l}
\toprule
 & OCTANE \cite{dhaun2014optical} & Diurnal Variation \cite{dhaun2014optical} & Normative & i-Test \cite{dhaun2014optical} & Prevent Dementia \cite{ritchie2012prevent} & GCU Topcon \cite{moukaddem2022comparison} & Total \\
  \midrule
\multicolumn{1}{l}{Subjects} & 46 & 20 & 1 & 21 & 121 & 24 & 233\\
\multicolumn{1}{r}{Control/Case} & 0 / 46 & 20 / 0 & 1 / 0 & 11 / 10 & 56 / 65 & 24 / 0 & 112 / 121 \\
\multicolumn{1}{r}{Male/Female} & 24 / 22 & 11 / 9 & 1 / 0 & 0 / 21 & 66 / 55 & 14 / 9 & 116 / 116 \\
\multicolumn{1}{r}{Right/Left eyes} & 46 / 0 & 20 / 0 & 1 / 1 & 21 / 21 & 117 / 115 & 22 / 21 & 227 / 158 \\
\multicolumn{1}{r}{Age (mean (SD))} & 47.5 (12.3) & 21.4 (2.3) & 23.0 (0.0) & 32.8 (5.4) & 50.8 (5.6) & 21.8 (7.9) & 42.9 (13.7) \\
\multicolumn{1}{l}{Device manufacturer} & Heidelberg & Heidelberg & Heidelberg & Heidelberg & Heidelberg & Topcon & All \\
\multicolumn{1}{l}{Device type} & Standard & Standard & FLEX & FLEX & Standard & DRI Triton Plus & All \\
\multicolumn{1}{l}{nEDI / EDI} & EDI & EDI & Both & EDI & Both &  &  \\
\multicolumn{1}{l}{Average ART} & 100 & 100 & 9 & 50 & 100 &  &  \\
\multicolumn{1}{l}{Scan location} &  &  &  &  & & & \\
\multicolumn{1}{r}{Horizontal/Vertical} & 168 / 0 & 55 / 50 & 4 / 4 & 76 / 76 & 381 / 369 & 132 / 139 & 816 / 638 \\
\multicolumn{1}{r}{Volume/Radial/Peripapillary} & 0 / 0 / 0 & 0 / 0 / 66 & 365 / 0 / 0 & 2,408 / 0 / 0 & 0 / 0 / 0 & 0 / 1,307 / 0 & 2,773 / 1,307 / 66 \\
\multicolumn{1}{r}{Total B-scans} & 168 & 171 & 373 & 2,560 & 750 & 1,578 & 5,600\\
\bottomrule
\end{tabular}}}}
\caption{Image characteristics of the cohorts used for building Choroidalyzer \cite{engelmann2023choroidalyzer}.}
\label{supptab:pop_choroidalyzer} 
\end{table}

\begin{table}[tbh]
\centering
{\small
\scalebox{0.75}{\begin{tabular}{@{}lccc|c@{}}
\toprule
\hspace{3em}                    & \hspace{1em}OCTANE\hspace{1.3em}       & \hspace{1em}i-Test\hspace{1.3em}     &\hspace{1em}Normative\hspace{1.3em}  & Total        \\ \midrule
Subjects            & 47           & 5          & 30         & 82           \\
Male/Female         & 24 / 23     & 0 / 5      & 20 / 10    & 44 / 38     \\
Right/Left eyes     & 47 / 0       & 5 / 5      & 29 / 29    & 81 / 34      \\
Age (mean (SD))     & 48.8 (12.9) & 34.4 (3.4) & 49.1 (7.0) & 48.0 (11.2) \\
HRA+OCT Module             & Standard            & FLEX          & Standard          &    Both          \\
Horizontal/Vertical scans & 166 / 0      & 16 / 16    & 57 / 54    & 239 / 70     \\
Volume scans           & 174          & 186        & 46         & 406          \\
Total B-scans         & 340          & 218        & 157        & 715          \\ \bottomrule
\end{tabular}}
}
\caption{Image characteristics of the cohorts used for building DeepGPET \cite{burke2023open}.}
\label{supptab:pop_deepgpet} 
\end{table}

\subsubsection{SLO Segmentation}
The following datasets were used for SLO segmentation: RAVIR \cite{hatamizadeh2022ravir}, an open source dataset of SLO images with varying degrees of retinal pathology. Normative \cite{cameron2016modulation}, a small dataset of SLO images collected in-house. SLO images captured from the PREVENT Dementia \cite{ritchie2012prevent, ritchie2013prevent} and i-Test \cite{dhaun2014optical} cohorts were also used (285 eyes and 186 eyes, respectively) and were described qualitatively above. FutureMS \cite{chen2022longitudinal, kearns2022futurems}, a cohort of individuals with newly diagnosed relapsing-remitting multiple sclerosis (MS) (15 eyes). More details can be found here \cite{burke2024sloctolyzer}.

\begin{table}[tbh]
\centering
{\small

\scalebox{0.75}{\centerline{\begin{tabular}{llllp{1.75cm}p{1.75cm}p{2.25cm}l}
\toprule
Study & Participants & Images & Right eyes & Retinal pathology & HRA+OCT Module & Image resolution, pixels & Location \\
\midrule
RAVIR \cite{hatamizadeh2022ravir} & 23 & 23 & 14 & Yes & Standard & 768 $\times$ 768 & Disc \\
Healthy \cite{cameron2016modulation} & 7 & 7 & 7 & No & Standard & 1536 $\times$ 1536 & Disc \\
PREVENT \cite{ritchie2012prevent, ritchie2013prevent} & 144 & 285 & 142 & No & Standard & 1536 $\times$ 1536 & Disc \\
i-Test \cite{dhaun2014optical} & 93 & 186 & 93 & No & Standard \& FLEX & 768 $\times$ 768 & Macula \\
FutureMS \cite{kearns2022futurems, chen2022longitudinal} & 15 & 15 & 9 & No & Standard & 1536 $\times$ 1536 & Disc \\
\bottomrule
\end{tabular}}}}
\caption{Image characteristics of the five cohorts used to build SLOcolyzer's segmentation module \cite{burke2024sloctolyzer}. Image resolution is in pixels (for both lateral and axial directions), location refers to the centring of the scan, i.e. if it's macula-centred disc-centred.}
\label{supptab:pop_sloctolyzer} 
\end{table}


\newpage

\subsection{DeepGPET's robustness to peripapillary choroids}
\begin{figure}[tbh]
    \centering
    \includegraphics[width=\textwidth]{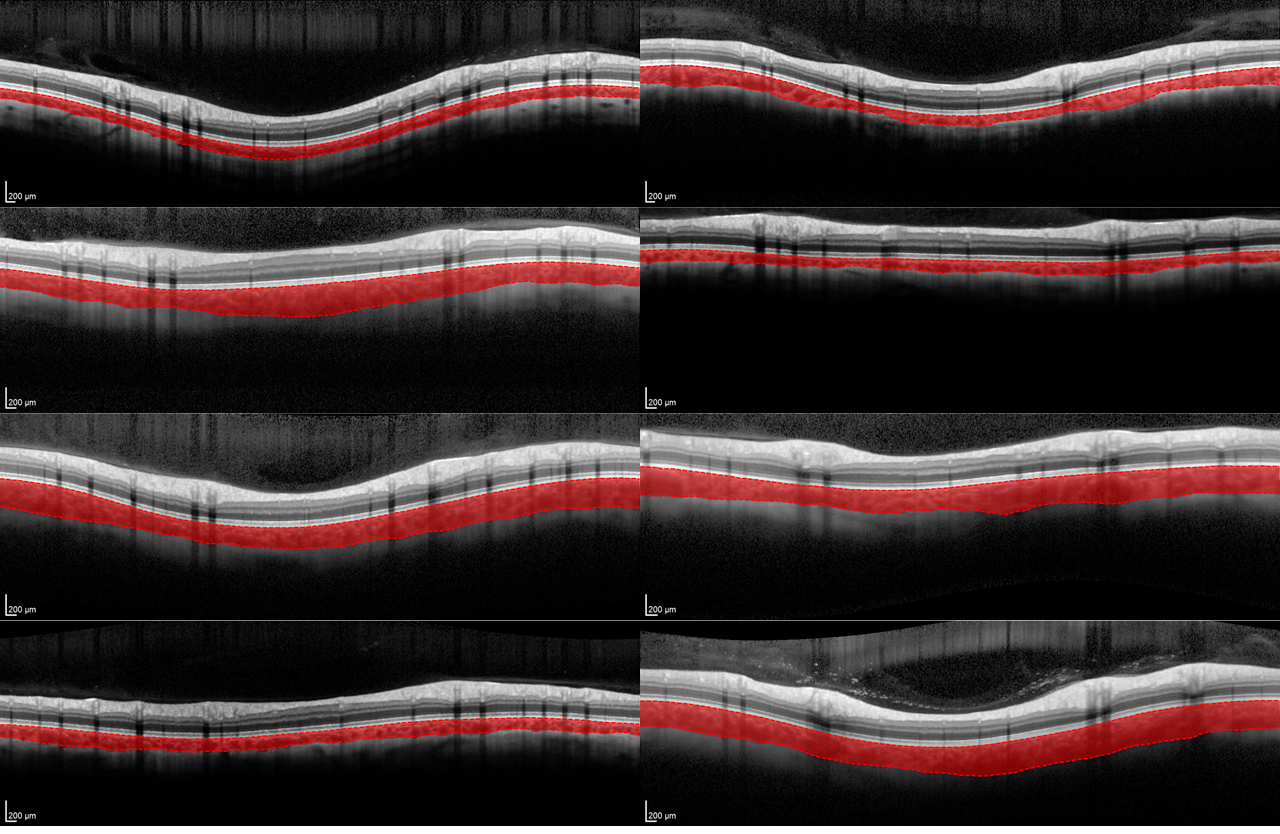}
    \caption{A selection of peripapillary choroids from 8 different eyes from the DVCKD cohort \cite{dhaun2014optical}, with successful choroid segmentation after application of DeepGPET \cite{burke2023open}.}
    \label{suppfig:deepgpet_peripapillary}
\end{figure}

Supplementary \cref{suppfig:deepgpet_peripapillary} shows successful choroid segmentations after application of DeepGPET \cite{burke2023open}, a model which was trained only on macula-centred OCT B-scans. The choroids were selected to show DeepGPET's robustness to image quality, choroid size and extent of sinuosity for different peripapillary OCT B-scan. The peripapillary choroids shown here were sourced from the DVCKD cohort \cite{dhaun2014optical}.

\newpage

\subsection{Layer definition}
Supplementary \cref{supptab:layers_computed} defines the different layers of the retina and choroid which OCTolyzer is capable of making measurements on for the three OCT data types which OCTolyzer supports.
\begin{table}[tbh]
\centering
{\small
\scalebox{0.99}{\begin{tabular}{ll}
\toprule
Abbreviation & Layer \\
\midrule
ILM -- RNFL & Retinal Nerve Fiber Layer \\
RNFL -- GCL & Ganglion Cell Layer \\
GCL -- IPL & Inner Plexiform Layer \\
IPL -- INL & Inner Nuclear Layer \\
INL -- OPL & Outer Plexiform Layer \\
OPL -- ELM & External Limiting Membrane \\
ELM -- PR1 & Photoreceptor Layer 1 \\
PR1 -- PR2 & Photoreceptor Layer 2 \\
PR2 -- RPE & Retinal Pigment Epithelium \\
RPE -- BM & Bruch's Membrane Complex \\
ILM -- ELM & Inner retinal layers \\
ELM -- BM & Outer retinal layers \\
ILM -- BM & All retinal layers \\
BM -- CHOR & Choroid\\
\bottomrule
\end{tabular}}}
\caption{All available layers which OCTolyzer measures. ILM, inner limiting membrane.}
\label{supptab:layers_computed} 
\end{table}


\newpage

\subsection{Detailed thickness map diagram}
\begin{figure}[tbh]
    \centering
    \includegraphics[width=0.99\textwidth]{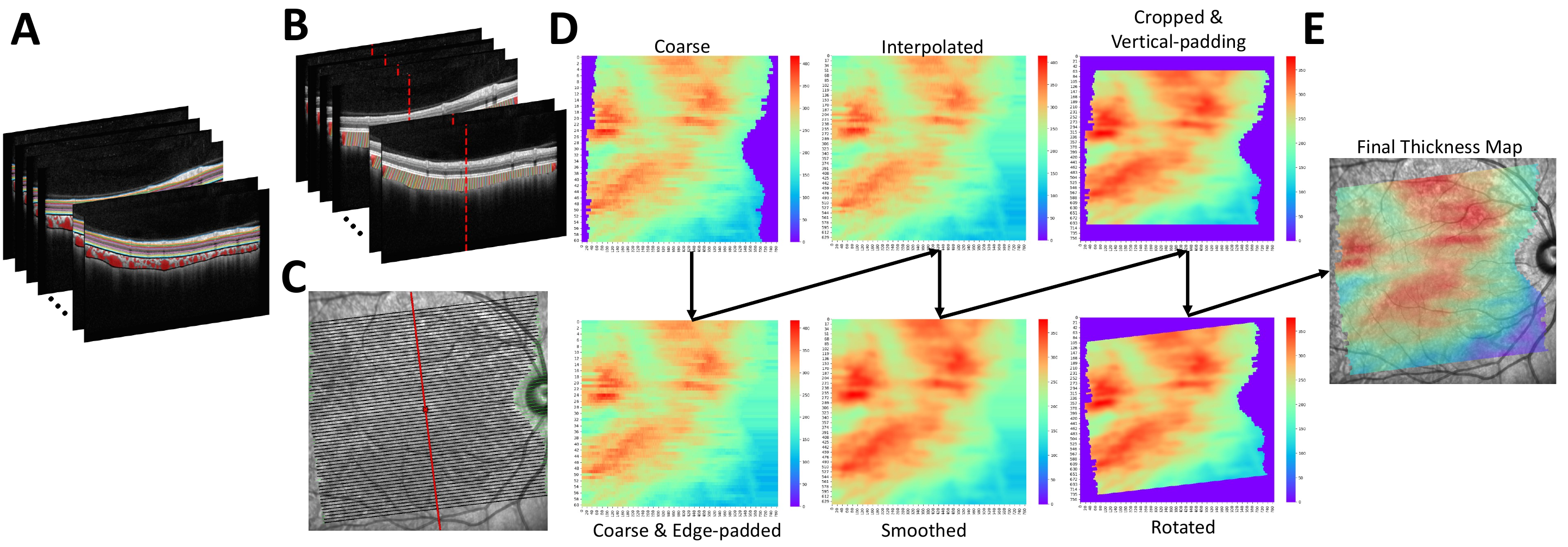}
    \caption{Detailed diagram on generating thickness maps, using the choroidal layer as an example. (A) OCT volume of sequential B-scans, with retinal and choroidal layer segmentations overlaid. (B) Valid thickness measurements taken of the choroid across each layer segmentation in every OCT B-scan (C) SLO image with lines of acquisition for each B-scan (green), with the distance the layer segmentation reached per B-scan (black), with the horizontal position on the B-scan of the fovea overlaid as a red line. (D) Step-by-step process of generating the thickness map. (E) The final choroid thickness map overlaid onto the SLO image.}
    \label{suppfig:thickness_map_diagram}
\end{figure}

Supplementary \cref{suppfig:thickness_map_diagram} shows a detailed diagram of how the thickness maps are generated from the segmentations of an OCT volume. The full, detailed pipeline is described below.

All valid thicknesses are measured for each B-scan (panel B, multicoloured), and each B-scans' thickness array is aligned with the fovea from the localiser SLO (panel B, red dotted lines). thickness array alignment per B-scan is required since the anatomical layer segmentations (panel C, black) do not necessarily cover the lateral width of the acquisition region of interest (panel C, black-on-green).

The aligned thickness arrays then follow this step-by-step process (panel D): the coarse thickness map is generated for each B-scan by vertically stacking the aligned thickness arrays. This is padded horizontally by duplicating the edge values. This is then interpolated to the same pixel resolution as the SLO localiser using bi-linear interpolation. A Gaussian filter whose standard deviation is the pixel distance between the parallel, OCT B-scans is used for smoothing. This is then vertically padded to centre the map onto the fovea of the localiser SLO, and the thickness values outside of the ends the original B-scan segmentations are set to 0 (panel C, black). Finally, this map is rotated to the angle of elevation of the region of interest (panel C, green). 

\newpage

\subsection{Overlap index demonstration}

\begin{figure}[tbh]
    \centering
    \includegraphics[width=\textwidth]{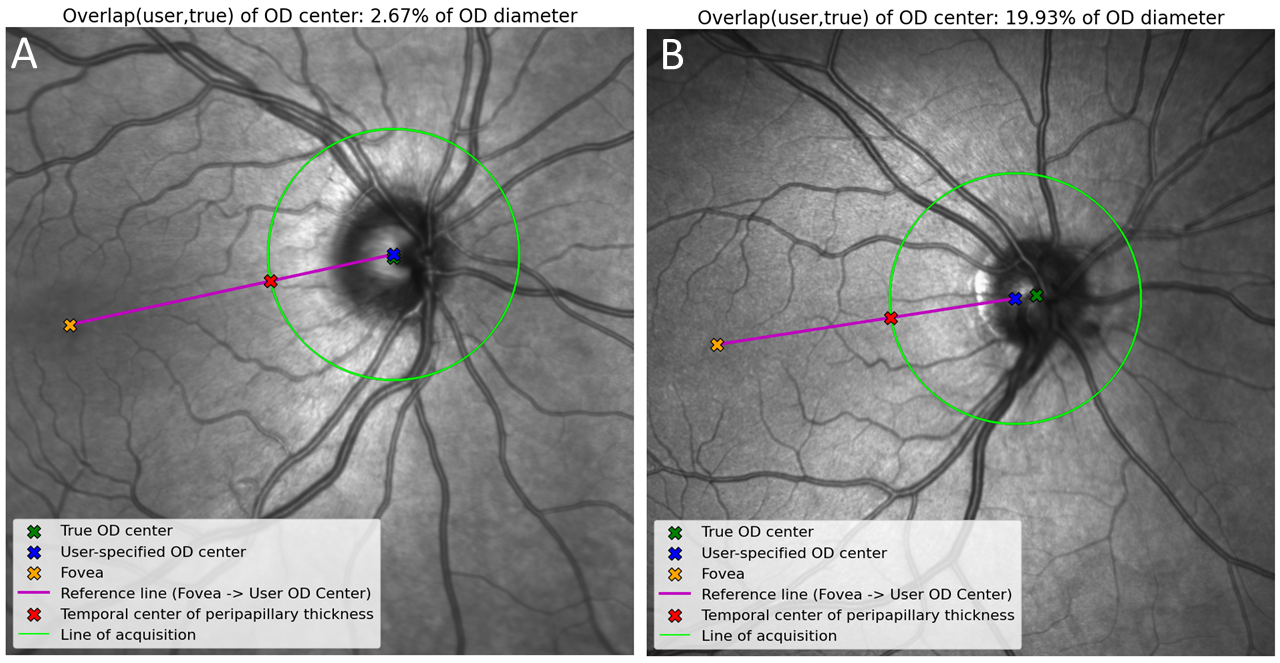}
    \caption{Overlap index measured and visualised for two OCT peripapillary scans. The images are (optionally) saved out during processing of an OCT peripapillary B-scan to visualise the overlap, as well as the alignment of the thickness profiles to the orientation of the fovea and optic disc centre. (A) An example where the overlap index is within the 15\% threshold, where the blue cross (the centre of the acquisition's B-scan) and green cross (the centre of the optic disc, measured with the SLO analysis suite), are within 3\% of the optic disc diameter. (B) an example where the overlap index exceeds the 15\% threshold. Here, the acquisition line-scan is clearly off-centre from the centre of the optic disc. OD, optic disc.}
    \label{suppfig:overlap_index_peri}
\end{figure}

\newpage

\subsection{Metadata output from OCTolyzer}

\begin{table}[tbh]
\centering
{\small
\scalebox{0.7}{\centerline{\begin{tabular}{l|l}
\toprule
Key & Description \\
\midrule
Filename & Filename of the SLO+OCT file analyse. \\
FAILED & Boolean flag on whether file unexpectedly failed to be analysed. \\
& \\
eye & Type of eye, either Right or Left. \\
Bscan\_type & Type of OCT scan acquired. One of H(orizontal)-line, V(ertical)-line;A(rtery)V(ein)-line, P(osterior)pole and Peripapillary. \\
Bscan\_resolution\_x & Number of columns of B-scan, typically 768 or 1536 for Heidelberg. \\
Bscan\_resolution\_y & Number of rows of B-scan, typically 768 or 496 for Heidelberg. \\
Bscan\_scale\_z & Micron distance between successive B-scans in a Posterior pole acquisition. Is 0 for all other Bscan\_types. \\
Bscan\_scale\_x & Pixel lengthscale in the horizontal direction B-scan/SLO, measured in microns per pixel. \\
Bscan\_scale\_y & Pixel lengthscale in the vertical direction in the B-scan, measured in microns per pixel. \\
bscan\_ROI\_mm & Region of interest (distance) captured by each B-scan measured in mm. \\
& \\
scale\_units & Units of the lengthscales, this is fixed as microns per pixel. \\
avg\_quality & Heidelberg-provided signal-to-noise ratio of the B-scan(s). \\
retinal\_layers\_N & Number of retinal layer segmentations extracted from metadata. \\
scan\_focus & Scan focus of the acquisition, in Dioptres. This decides the scaling and is a gross measure of refractive error. \\
visit\_date & Date of acquisition. \\
exam\_time & Time of acquisition. \\
& \\
slo\_resolution\_px & Number of rows/columns in the square-shaped SLO image (typically 768 or 1536). \\
field\_of\_view\_mm & Field of view captured during acquisition, usually between 8 and 9 mm if field size is 30 degrees. \\
slo\_scale\_xy & Pixel lengthscale of the SLO image, and is typically the same for both directions. \\
location & Whether scan is macula-centred or disc-centred. Is either ``macular'' or ``peripapillary'' \\
field\_size\_degrees & Field of view in degrees, typically 30. \\
slo\_modality & Modality used for SLO image capture. OCTolyzer supports grayscale NIR cSLO images currently.\\
acquisition\_angle\_degrees & Angle of elevation from horizontal image axis of acquisition for Posterior pole scans. \\
& \\
Bscan\_fovea\_x & Horizontal pixel position of the fovea on the OCT B-scan (if visible in one of the scans, only relevant for macular OCT). \\
Bscan\_fovea\_y & Vertical pixel position of the fovea on the OCT B-scan (if visible in one of the scans, only relevant for macular OCT). \\
slo\_fovea\_x & Horizontal pixel position of the fovea on the SLO image, if visible. \\
slo\_fovea\_y & Vertical pixel position of the fovea on the SLO image, if visible. \\
slo\_missing\_fovea & Boolean value flagging whether fovea is missing from data (either due to acquisition or segmentation failure). \\
& \\
optic\_disc\_overlap\_index\_\% & \% of the optic disc diameter, defining how off-centre a peripapillary image acquisition is from the optic disc centre.\\
optic\_disc\_overlap\_warning & Boolean value, flagging if the overlap index is greater than 15\%, the empirical cut-off to warn end-user of an off-centre scan. \\
optic\_disc\_x & Horizontal pixel position of the optic disc centre on the SLO image, if visible.\\
optic\_disc\_y & Vertical pixel position of the optic disc centre on the SLO image, if visible.\\
optic\_disc\_radius\_px & Pixel radius of the optic disc. \\
& \\
thickness\_units & Units of measurement for thickness, always in $\mu$m (microns). \\
vascular\_index\_units & Units of measurement for choroid vascular index, always dimensionless (no units, but is a ratio between 0 and 1). \\
vessel\_density\_units & Units of measurement for choroid vessel density, always in $\mu$m$^2$ (square microns).\\
area\_units & Units of measurements for area, always in mm$^2$ (square millimetres). \\
volume\_units & Units of measurements for volume, always in mm$^3$ (cubic millimetres). \\
linescan\_area\_ROI\_microns & For single-line, macular OCT, this is the micron distance defining the fovea-centred region of interest. \\
choroid\_measure\_type & Whether the choroid is measured column-wise (per A-scan) or perpendicularly. Always per A-scan for peripapillary OCT. \\
& \\
acquisition\_radius\_px & Pixel radius of the acquisition line around the optic disc for peripapillary OCT. \\
acquisition\_radius\_mm & Millimetre radius of the acquisition line around the optic disc for peripapillary OCT. \\
acquisition\_optic\_disc\_center\_x & Horizontal pixel position of the optic disc centre, as selected by the user during peripapillary OCT acquisition.\\
acquisition\_optic\_disc\_center\_y & Vertical pixel position of the optic disc centre, as selected by the user during peripapillary OCT acquisition. \\
\bottomrule
\end{tabular}}}
}
\caption{Metadata extracted and inferred using OCTolyzer when processing a ``.vol'' RAW export file from a Heidelberg Engineering imaging device.}
\label{supptab:metadata_keys} 
\end{table}

\newpage

\subsection{OCTolyzer's interface}

\begin{figure}[tbh]
    \centering
    \includegraphics[width=\textwidth]{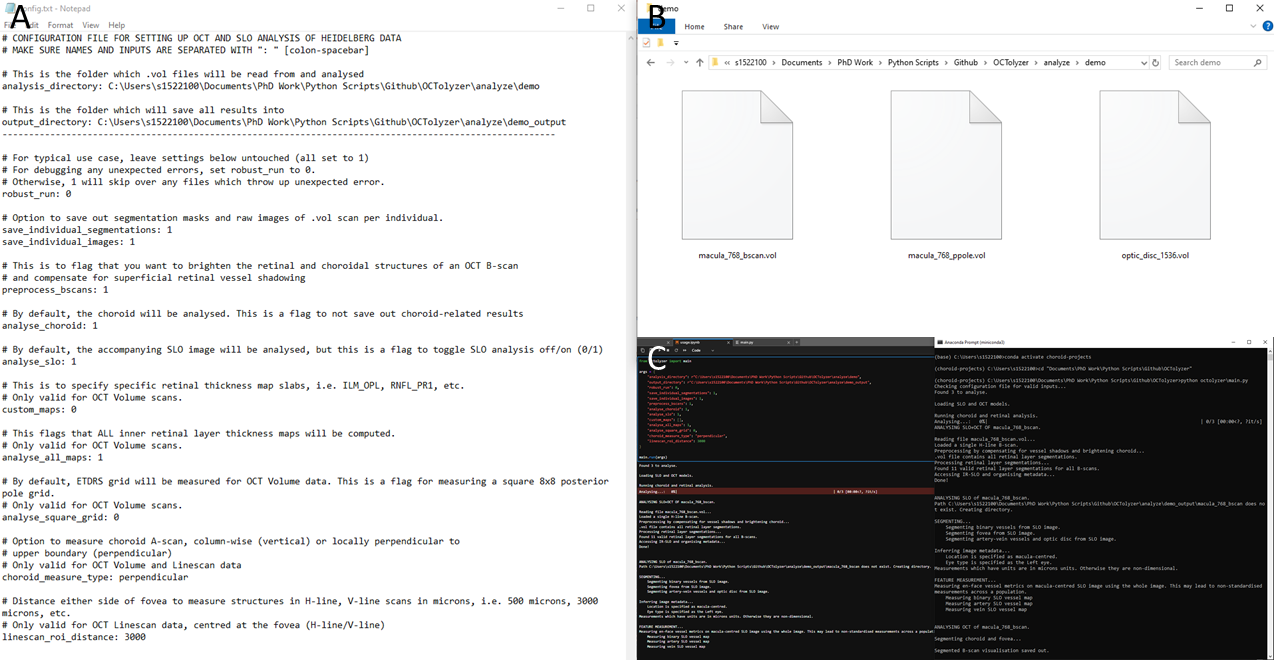}
    \caption{Demonstration of setting up and running OCTolyzer on a batch of demonstrative data. (A) configuration file with user-specified inputs. (B) The folder storing the files (all ``.vol'' in this example) to be analysed. (C) Running OCTolyzer on a batch of data can be done from a python integrated development environment (left) or via the terminal (right).}
    \label{suppfig:octolyzer_input}
\end{figure}

\begin{figure}[tb]
    \centering
    \includegraphics[width=\textwidth]{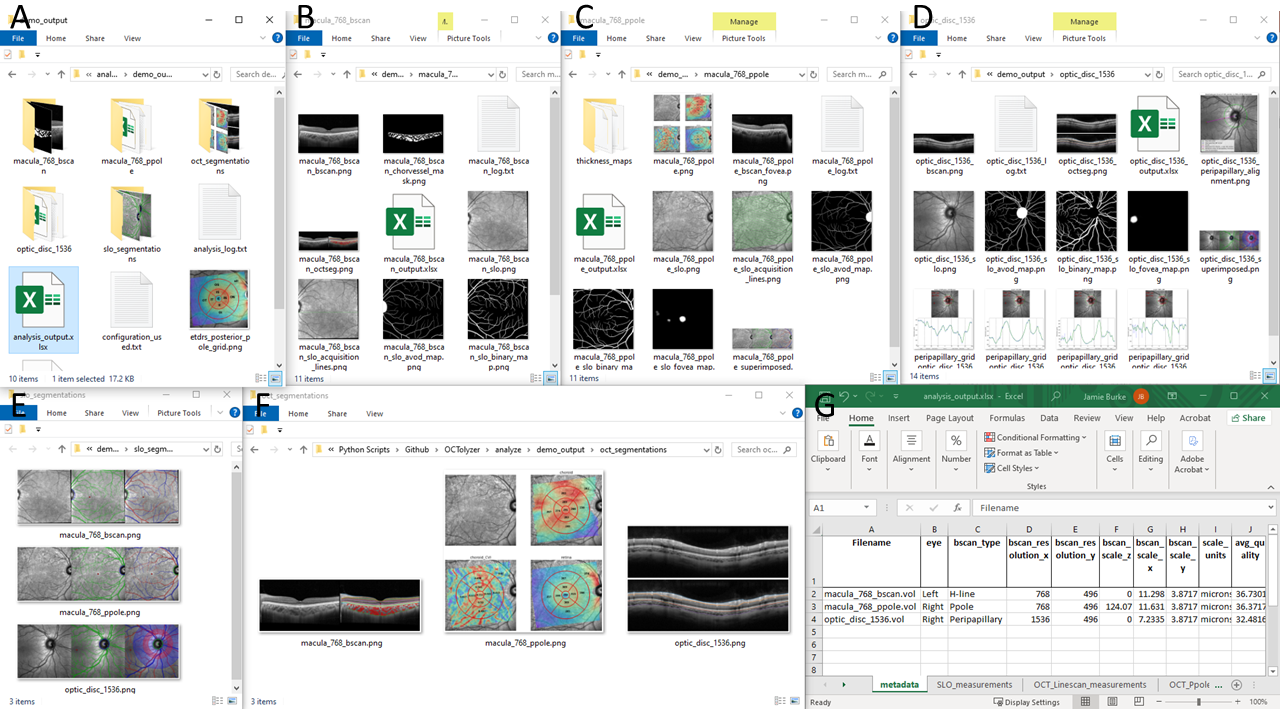}
    \caption{Output files and folders from running OCTolyzer on a batch of three files (all ``.vol'' in this example), each a different OCT data type which is supported. (A) The folder storing all outputs from batch processing, including folders with results for each individually processed file, a process log, the configuration used and summary measurement and metadata output. (B -- D) Exemplar output for each of the OCT data types which OCTolyzer supports: single macular B-scan, macular volume and peripapillary B-scan. (E -- F) Composite segmentations of the SLO and OCT segmentation masks for real-time segmentation quality inspection. (G) Summary output file storing key metadata and measurements from OCTolyzer's measurement module for the SLO and OCT data.}
    \label{suppfig:octolyzer_output}
\end{figure}

\end{document}